\PassOptionsToPackage{sort&compress}{NJDnatbib}
\documentclass[ASNA,twocolumn]{USG}

\newcommand{\kunit}{\ensuremath{\mathrm{W\,m^{-1}\,K^{-1}}}}

\newcommand{\chem}[1]{\ensuremath{\mathrm{#1}}}

\setcitestyle{numbers,square}

\newenvironment{fixedfigure}
  {\begin{figure*}[!tp]\centering}
  {\end{figure*}}

\newenvironment{fixedtable}
  {\begin{table*}[!tp]\centering}
  {\end{table*}}

\allowdisplaybreaks[1]

\graphicspath{{./images/}}
\hypersetup{pdftitle={High-Throughput Identification of Ultralow-Thermal-Conductivity Crystal Phases},pdfauthor={Xiao-Qi Han; Ze-Feng Gao; Ke-Xin Pang; Peng-Jie Guo; Zhong-Yi Lu},colorlinks=true,linkcolor=blue,citecolor=blue,urlcolor=blue}

\articletype{RESEARCH ARTICLE}
\journal{Manuscript draft}
\volume{}
\copyyear{2026}
\startpage{1}
\received{}
\revised{}
\accepted{}
\articledoi{}
\copyright{Draft manuscript.}

\begin{document}
\title{Accelerated Discovery of Ultralow-Thermal-Conductivity Crystals via Structure-to-Property Prediction and Machine-Learning Interatomic Potentials}
\author[1,2]{Xiao-Qi Han*}
\author[1,2]{Ze-Feng Gao*}
\author[3]{Ke-Xin Pang}
\author[1,2]{Peng-Jie Guo}
\author[1,2]{Zhong-Yi Lu}
\address[1]{School of Physics, Renmin University of China, Beijing, 100872, P.R. China}
\address[2]{AI-Driven Quantum Materials Research Center, School of Physics, Renmin University of China, Beijing, 100872, P.R. China}
\address[3]{State Key Laboratory of Metastable Materials Science and Technology and Hebei Key Laboratory of Microstructural Material Physics, School of Science, Yanshan University, Qinhuangdao 066004, China}
\corres{Ze-Feng Gao and Zhong-Yi Lu. Email: \email{zfgao@ruc.edu.cn}; \email{zlu@ruc.edu.cn}.}

\begin{abstract}{
Identifying crystals with ultralow lattice thermal conductivity is important for thermoelectric energy conversion and thermal management. However, the high computational cost of first-principles phonon-transport calculations limits large-scale screening across extensive materials spaces. Here, we develop a multilevel high-throughput screening framework that integrates structure-to-property prediction, machine-learning interatomic potentials, and first-principles calculations to accelerate the discovery of crystals with ultralow lattice thermal conductivity. We apply this framework to 1,953,259 crystal structures from the Materials Project and Crystoria databases. Using our proposed DANA-$\kappa$ (Distribution-Aware Neighborhood Attention) model, we identify 656,693 candidates with predicted lattice thermal conductivities below $2~\mathrm{W\,m^{-1}\,K^{-1}}$. Subsequent structural, electronic, and thermodynamic filtering, followed by GRACE-based phonon-transport calculations, yields 73 candidates for DFT validation for which no lattice thermal conductivity calculations have been reported in the existing literature. Among the 59 dynamically stable materials for which DFT thermal-conductivity calculations are completed, 58 exhibit lattice thermal conductivities below $2~\mathrm{W\,m^{-1}\,K^{-1}}$, including 29 below $0.5~\mathrm{W\,m^{-1}\,K^{-1}}$. The lowest value is found in $\mathrm{CsKAu_2}$, at $0.1826~\mathrm{W\,m^{-1}\,K^{-1}}$. DANA-$\kappa$ is built on a coupled atom-graph and bond-angle line-graph architecture, incorporating two-pass neighborhood slot attention and distribution-aware aggregation to enhance the representation of local atomic environments. For the 58 DFT-validated materials, the model achieves a logarithmic mean absolute error (MAE) of 0.527, with 72.9\% of its predictions falling within a factor of two of the DFT values. Across random, space-group-disjoint, and low-thermal-conductivity out-of-distribution test sets, it achieves an average MAE of 0.677, outperforming the ALiEGNN baseline. DANA-$\kappa$ requires an average inference time of only 3.62 ms per structure, compared with approximately 15.9 min for GRACE-based phonon-transport calculations and 24.4 h for DFT calculations per structure. This multilevel high-throughput framework bridges large-scale crystal-structure screening and first-principles validation, providing an efficient and scalable strategy for discovering crystals with ultralow lattice thermal conductivity.}
\end{abstract}
\maketitle

\section{Introduction}\label{sec:introduction}

Materials with low lattice thermal conductivity are important for thermoelectric energy conversion, thermal insulation, and heat-flow control, because suppressing phonon-mediated conduction reduces parasitic heat leakage and expands the design space for thermal functionality~\cite{Snyder2008,Qian2021}. Representative systems such as SnSe and full-Heusler semiconductors demonstrate that ultralow conductivity can occur in crystals with different chemistries and levels of structural complexity~\cite{Zhao2014,He2016}. Identifying such materials across large crystal spaces remains constrained by the cost of reliable transport calculations. First-principles predictions require relaxed structures, harmonic and anharmonic interatomic force constants, and Brillouin-zone integration of the transport equations. The numerous supercell force evaluations and convergence checks make this workflow demanding at million-structure scale~\cite{Phono3py2023,Phonix2026}. Moreover, particle-like and coherence-mediated heat transport can coexist, making a consistent transport definition and independent physical validation essential for comparing low-conductivity candidates~\cite{Simoncelli2019,Simoncelli2022}.

Universal machine-learning interatomic potentials (uMLIPs) learn atomic interactions from first-principles data, replacing repeated electronic-structure calculations with efficient energy and force evaluations and substantially reducing DFT cost. Representative models include M3GNet~\cite{M3GNet2022}, CHGNet~\cite{CHGNet2023}, MACE~\cite{MACEFoundation2025}, and more recent eSEN~\cite{ESEN2025}, Orb-v3~\cite{OrbV32025}, GRACE~\cite{GRACE2026}, and DPA4~\cite{DPA42026}. For lattice thermal conductivity, Póta et al. evaluated several foundation potentials combined with the Wigner transport equation across 103 solids~\cite{Pota2026}. MACE-MP-0~\cite{MACEFoundation2025} predicted thermal conductivities within a factor of two of DFT for approximately 69\% of the materials; for BeO, its zero-shot error was about 7\% at more than three orders of magnitude lower cost, and fine-tuning reduced the LiBr error to approximately 2\%. These results support uMLIPs for high-throughput transport screening at substantially reduced cost.

In parallel, direct structure-to-property prediction offers an alternative route for high-throughput screening: surrogate models map structural representations to thermal conductivity, bypassing explicit force evaluation and phonon-transport solution. Graph networks such as CGCNN~\cite{CGCNN2018}, MEGNet~\cite{MEGNet2019}, ALIGNN~\cite{ALIGNN2021} and ALiEGNN~\cite{Phonix2026} provide general frameworks for learning composition, local geometry, and bond-angle information, while expanding anharmonic databases have enabled conductivity-specific models~\cite{Phonix2026}. Wang and co-workers compared 15 methods on 6,966 Phonix-derived entries using random, space-group, and low-conductivity out-of-distribution (OOD) splits, highlighting the importance of evaluating both average prediction accuracy and generalization to unseen structural distributions~\cite{Surrogate2026}. Despite these advances, identifying ultralow-thermal-conductivity materials across large crystal databases remains challenging.

Here, we develop a multilevel high-throughput screening framework
that integrates structure-to-property prediction, machine-learning
interatomic potentials, and first-principles calculations to
accelerate the discovery of crystals with ultralow lattice thermal
conductivity. We apply this framework to crystal structures from
the Materials Project and Crystoria databases, identifying 58
previously unreported low-thermal-conductivity candidates with
DFT-calculated $\kappa_L < 2~\mathrm{W\,m^{-1}\,K^{-1}}$,
among which $\mathrm{CsKAu_2}$ exhibits the lowest lattice thermal
conductivity. We further investigate the phonon transport
mechanisms of representative compounds to elucidate the
microscopic origins of their low thermal conductivity. The framework incorporates our proposed DANA-$\kappa$
(Distribution-Aware Neighborhood Attention), a direct
structure-to-property model for predicting lattice thermal
conductivity. DANA-$\kappa$ outperforms the ALiEGNN baseline
in predictive accuracy across random, space-group-disjoint,
and low-conductivity out-of-distribution test sets, while
requiring only 3.62 ms per structure for inference, compared
with approximately 15.9 min for GRACE-based phonon-transport
calculations~\cite{GRACE2026}. In addition, our extended benchmark of interatomic
potentials demonstrates that GRACE achieves the highest accuracy
in lattice thermal conductivity calculations among the evaluated
potentials. Finally, we examine the coexistence of low lattice
thermal conductivity and nontrivial crystalline topology in BaPd.
This work establishes an integrated computational strategy
for accelerating the discovery and physical characterization
of ultralow-thermal-conductivity crystals.

\section{Results and discussion}\label{sec:results}

\begin{fixedfigure}
\centering
\includegraphics[width=\textwidth,height=.69\textheight,keepaspectratio]{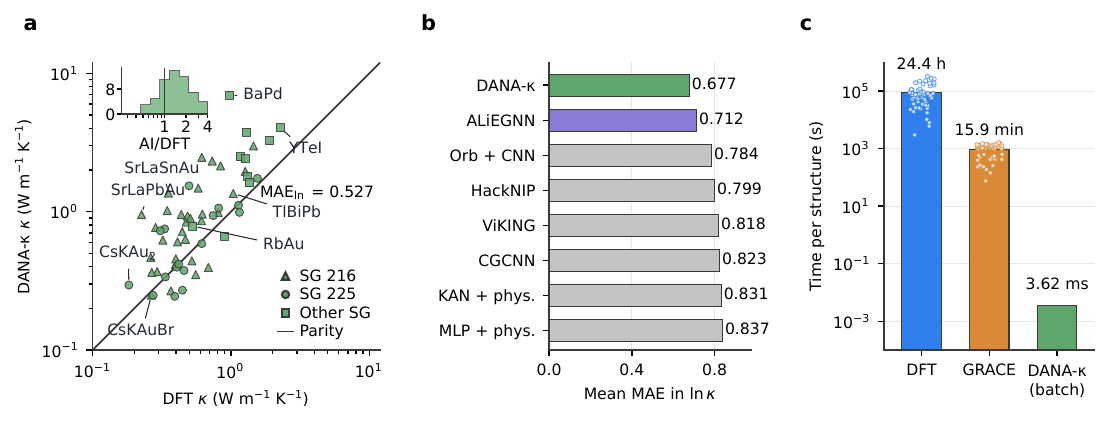}
\caption{\textbf{Prediction accuracy and computational cost in ultralow-thermal-conductivity screening.}
\textbf{a,} DFT total lattice thermal conductivities at 300~K versus predictions from the trained DANA-$\kappa$ model for 58 structures with completed DFT thermal-conductivity calculations. The solid line indicates parity. Triangles, circles, and squares denote space groups 216, 225, and all other space groups, respectively. Selected compounds are annotated. The inset shows data points with prediction-to-DFT ratios between 0.25 and 4.
\textbf{b,} The eight best-performing models ranked by the mean MAE in natural-logarithmic thermal conductivity, averaged equally across the Random, space-group, and low-conductivity OOD splits. DANA-$\kappa$ represents the model developed in this work. Other values are taken from the literature~\cite{Surrogate2026}.
\textbf{c,} Observed computational wall time per structure. The DFT and GRACE bars show the mean wall time across the 58-structure cohort, with individual points representing the corresponding materials. The DANA-$\kappa$ bar shows the average inference time over the full dataset, amounting to 3.62~ms per source record.}
\label{fig:accuracy-efficiency}
\end{fixedfigure}

\subsection{High-throughput screening and identification of ultralow-thermal-conductivity materials}\label{sec:dft}

Our multilevel screening workflow, combining DANA-\(\kappa\), the universal machine learning interatomic potentials GRACE~\cite{GRACE2026}, and DFT, identified a large number of semiconductor candidates with ultralow thermal conductivity. Among the 59 structures for which DFT thermal conductivity calculations were completed, 58 exhibited lattice thermal conductivities below \(2\ \mathrm{W\,m^{-1}\,K^{-1}}\), including 29 below \(0.5\ \mathrm{W\,m^{-1}\,K^{-1}}\). The lowest value was \(0.1826\ \mathrm{W\,m^{-1}\,K^{-1}}\) (Fig. ~\ref{fig:accuracy-efficiency}a).
As shown in Fig.~\ref{fig:architecture-screening}a,b, the initial dataset comprised 154,377 and 1,798,882 crystal structures from Materials Project~\cite{MaterialsProject2013} and Crystoria~\cite{pang2026crystorialargescalecrystaldataset}, respectively, totaling 1,953,259 structures. We first applied DANA-\(\kappa\) to predict the thermal conductivity of all structures and retained 656,693 candidates with predicted values of \(\hat{\kappa}_{\mathrm L}<2\ \mathrm{W\,m^{-1}\,K^{-1}}\). These candidates were then filtered according to structural, electronic, and thermodynamic criteria. Specifically, we restricted the primitive-cell size to no more than 20 atoms to limit the computational cost, required a band gap greater than 0.1 eV to exclude metallic and near-zero-gap systems, excluded magnetic structures, and imposed a hull-energy threshold of 100 meV/atom to remove high-energy metastable structures. This step yielded 47,588 candidates. An empirical computational-cost filter was used to define a shortlist of 4,520 structures; its cost proxies and selection thresholds are given in ppendix~\ref{app:computational-budget} and Table~\ref{tab:sm-cost-budget}. GRACE combined with Phono3py was used to evaluate their dynamical stability and thermal conductivity, yielding 3,410 structures that met the requirements for thermal-conductivity calculations. Among these, 1,712 had calculated thermal conductivities below \(2\ \mathrm{W\,m^{-1}\,K^{-1}}\). We subsequently performed DFT validation for 73 candidate materials not previously reported in the literature, of which 59 completed thermal-conductivity calculations. Ultimately, 58 structures exhibited DFT lattice thermal conductivities below \(2\ \mathrm{W\,m^{-1}\,K^{-1}}\), ranging from 0.1826 to \(1.9037\ \mathrm{W\,m^{-1}\,K^{-1}}\). Of these, 29 were below \(0.5\), 17 fell between 0.5 and \(1.0\), and 12 between 1.0 and \(2.0\ \mathrm{W\,m^{-1}\,K^{-1}}\). Phonon dispersions, temperature-dependent conductivities, group velocities, and three-phonon lifetimes for all 58 low-conductivity phases are provided in Appendix~\ref{app:all-transport}
and Figs.~\ref{fig:sm-transport-01}--\ref{fig:sm-transport-15}.

To evaluate the predictive capability of DANA-$\kappa$ (Distribution-Aware Neighborhood Attention model for lattice thermal conductivity prediction) for
screening materials with low thermal conductivity, we compared
its predictions against the DFT-calculated lattice thermal
conductivities of 58 structures. As shown in Fig.~\ref{fig:accuracy-efficiency}a and
Table~\ref{tab:dft-low-kappa}, DANA-$\kappa$ captures the variation in thermal
conductivity across a broad range of materials, achieving a
logarithmic mean absolute error ($\mathrm{MAE}_{\ln}$) of
0.527. The prediction errors are within a factor of two for
43 structures (72.9\%). Furthermore, among the 59 candidates
obtained from the final screening, 58 exhibit DFT-calculated
lattice thermal conductivities below
$2\ \mathrm{W\,m^{-1}\,K^{-1}}$. These results demonstrate
that the multilevel screening workflow effectively enriches
candidates with ultralow lattice thermal conductivity.

Fig.~\ref{fig:accuracy-efficiency}b and Table~\ref{tab:model-benchmark} further compare the predictive performance
of DANA-$\kappa$ with that of existing models across different
data splits. We use the equally weighted average of
$\mathrm{MAE}_{\ln}$ across three test sets---Random,
space-group, and low-conductivity OOD---as the overall
evaluation metric. DANA-$\kappa$ achieves an average error of
0.677, outperforming the baseline ALiEGNN model (0.711),
and demonstrates competitive predictive performance across
different evaluation settings.

Finally, Fig.~\ref{fig:accuracy-efficiency}c compares the computational costs of
DANA-$\kappa$, GRACE, and DFT in the actual screening workflow.
For the 58 structures subjected to subsequent calculations,
the average computational times per structure are approximately
24.4~h for DFT and 15.9~min for GRACE. In contrast,
DANA-$\kappa$ processes all 1,953,259 source records in a total
of 7,079.21~s, corresponding to an average of approximately
3.62~ms per record. This measurement includes model
initialization, CIF parsing, CPU-based graph construction,
data transfer, GPU inference, and database writing, thereby
reflecting the actual computational cost of the complete
batch-screening workflow. Although the hardware environments
and execution procedures differ among these computational
tasks, the results highlight the capability of DANA-$\kappa$
to process millions of crystal structures at low computational
cost. This efficiency enables rapid identification of
promising candidates from large-scale structural databases
and provides an efficient entry point for subsequent GRACE
calculations and DFT verification. Material-specific DFT computation times and their accounting definitions are given in Appendix~\ref{app:dft-methods} and Table~\ref{tab:sm-dft-times}.

\begin{fixedtable}
\small\setlength{\tabcolsep}{3pt}\renewcommand{\arraystretch}{1.04}
\caption{DANA-$\kappa$, GRACE and DFT lattice conductivities of all 58 crystal phases with $\kappa_{\rm DFT}<2$~\kunit{} at 300 K. Each five-column block is ordered by DFT conductivity; conductivity values are in \kunit.}
\label{tab:dft-low-kappa}
\begin{tabular*}{\textwidth}{@{\extracolsep{\fill}}lrrrr@{\hspace{12pt}}lrrrr@{}}
\toprule
\multicolumn{5}{c}{$\kappa_{\rm DFT}<0.5$} & \multicolumn{5}{c}{$0.5\leq\kappa_{\rm DFT}<2$} \\
\cmidrule(lr){1-5}\cmidrule(lr){6-10}
Formula & SG & DANA & GRACE & DFT & Formula & SG & DANA & GRACE & DFT \\
\midrule
$\mathrm{CsKAu_{2}}$ & 225 & 0.2949 & 0.1752 & 0.1826 & $\mathrm{CaInGaHg}$ & 216 & 0.8996 & 0.4647 & 0.5075 \\
$\mathrm{SrLaPbAu}$ & 216 & 0.9532 & 0.3983 & 0.2254 & $\mathrm{KBaHgBi}$ & 216 & 0.4409 & 0.4851 & 0.5209 \\
$\mathrm{CsBaHgSb}$ & 216 & 0.4670 & 0.3831 & 0.2626 & $\mathrm{RbAu}$ & 187 & 0.7862 & 0.4919 & 0.5295 \\
$\mathrm{CsKAuBr}$ & 216 & 0.2492 & 0.1824 & 0.2657 & $\mathrm{KBaCdBi}$ & 216 & 0.3511 & 0.5011 & 0.5553 \\
$\mathrm{CsKRbSb}$ & 216 & 0.3638 & 0.3996 & 0.2675 & $\mathrm{BaCaHgGe}$ & 216 & 1.4819 & 0.5202 & 0.5789 \\
$\mathrm{CsRbAu_{2}}$ & 225 & 0.2488 & 0.2307 & 0.2736 & $\mathrm{CaTlCdIn}$ & 216 & 0.8602 & 0.3543 & 0.6117 \\
$\mathrm{KNaHgTe}$ & 216 & 0.7703 & 0.3562 & 0.2849 & $\mathrm{CsRbCl_{2}}$ & 225 & 0.5874 & 0.7282 & 0.6143 \\
$\mathrm{CsKNaSb}$ & 216 & 0.3686 & 0.4339 & 0.2925 & $\mathrm{LuMgSnAu}$ & 216 & 2.4684 & 0.5133 & 0.6145 \\
$\mathrm{Rb_{2}HgS}$ & 225 & 0.7270 & 0.2167 & 0.3078 & $\mathrm{KCaHgSb}$ & 216 & 0.9615 & 0.5824 & 0.6195 \\
$\mathrm{NaTlHgTe}$ & 216 & 0.6223 & 0.3430 & 0.3212 & $\mathrm{KBaNaSn}$ & 216 & 0.3952 & 0.5505 & 0.6860 \\
$\mathrm{K_{2}HgSe}$ & 225 & 0.7505 & 0.3288 & 0.3309 & $\mathrm{CsKS}$ & 216 & 2.3229 & 1.2557 & 0.7313 \\
$\mathrm{Cs_{2}AuI}$ & 225 & 0.3376 & 0.1561 & 0.3350 & $\mathrm{KNa_{2}As}$ & 225 & 0.9372 & 0.7390 & 0.7428 \\
$\mathrm{CaLaPbAu}$ & 216 & 1.0178 & 0.5359 & 0.3456 & $\mathrm{NaLaBiAu}$ & 216 & 0.9871 & 0.6001 & 0.8086 \\
$\mathrm{SrLaSnAu}$ & 216 & 1.3628 & 0.4636 & 0.3530 & $\mathrm{Rb_{2}Pd}$ & 225 & 1.0611 & 1.1227 & 0.8137 \\
$\mathrm{CsRbAuCl}$ & 216 & 0.2674 & 0.1794 & 0.3675 & $\mathrm{BaBrCl}$ & 216 & 2.1440 & 1.0957 & 0.8395 \\
$\mathrm{Cs_{2}AuCl}$ & 225 & 0.2445 & 0.3452 & 0.3920 & $\mathrm{RbTlS}$ & 156 & 0.6623 & 0.7061 & 0.8947 \\
$\mathrm{RbSrHgBi}$ & 216 & 0.4119 & 0.3438 & 0.3953 & $\mathrm{BaPd}$ & 221 & 6.9368 & 1.5594 & 0.9805 \\
$\mathrm{RbBaHgSb}$ & 216 & 0.4562 & 0.6323 & 0.3986 & $\mathrm{TlBiPb}$ & 216 & 1.3556 & 0.2352 & 1.0370 \\
$\mathrm{KRbI_{2}}$ & 225 & 0.3971 & 0.4888 & 0.4064 & $\mathrm{CsAg}$ & 225 & 1.1128 & 1.7481 & 1.1321 \\
$\mathrm{RbSrHgSb}$ & 216 & 0.6051 & 0.5076 & 0.4102 & $\mathrm{KNa_{2}Bi}$ & 225 & 0.9891 & 1.3101 & 1.1494 \\
$\mathrm{Rb_{2}AuBr}$ & 225 & 0.4176 & 0.1869 & 0.4188 & $\mathrm{NaRbS}$ & 187 & 2.4876 & 1.0222 & 1.1616 \\
$\mathrm{BaNaHgSb}$ & 216 & 0.9529 & 0.5063 & 0.4368 & $\mathrm{RbBaBi}$ & 216 & 1.9531 & 0.8547 & 1.2662 \\
$\mathrm{NaHgAsPb}$ & 216 & 0.7163 & 0.2298 & 0.4437 & $\mathrm{CsAgS}$ & 187 & 2.4202 & 0.8967 & 1.2691 \\
$\mathrm{Cs_{2}AuBr}$ & 225 & 0.2705 & 0.2137 & 0.4451 & $\mathrm{CsAgSe}$ & 187 & 3.7507 & 1.2736 & 1.2892 \\
$\mathrm{Cs_{2}AgBr}$ & 225 & 0.3753 & 0.1803 & 0.4573 & $\mathrm{RbAgS}$ & 187 & 1.8099 & 1.1709 & 1.3198 \\
$\mathrm{KNaTlBi}$ & 216 & 0.6315 & 0.3800 & 0.4677 & $\mathrm{SrIBr}$ & 156 & 1.6212 & 1.4585 & 1.3594 \\
$\mathrm{MgTlBi}$ & 216 & 0.8394 & 0.4663 & 0.4739 & $\mathrm{RbBaAs}$ & 216 & 2.9884 & 1.1680 & 1.4524 \\
$\mathrm{NaSnHgSb}$ & 216 & 0.9297 & 0.2345 & 0.4851 & $\mathrm{Sr_{2}ZnSi}$ & 225 & 1.7409 & 1.9358 & 1.5517 \\
$\mathrm{PbCl_{2}}$ & 225 & 1.5372 & 0.4992 & 0.4967 & $\mathrm{BaTlP}$ & 156 & 3.2687 & 1.6639 & 1.9037 \\
\bottomrule\end{tabular*}
\par\vspace{5pt}
\end{fixedtable}

\subsection{Phonon transport mechanisms in representative ultralow-thermal-conductivity compounds}\label{sec:representative-transport}

\begin{fixedfigure}
\centering
\includegraphics[width=\textwidth,height=.69\textheight,keepaspectratio]{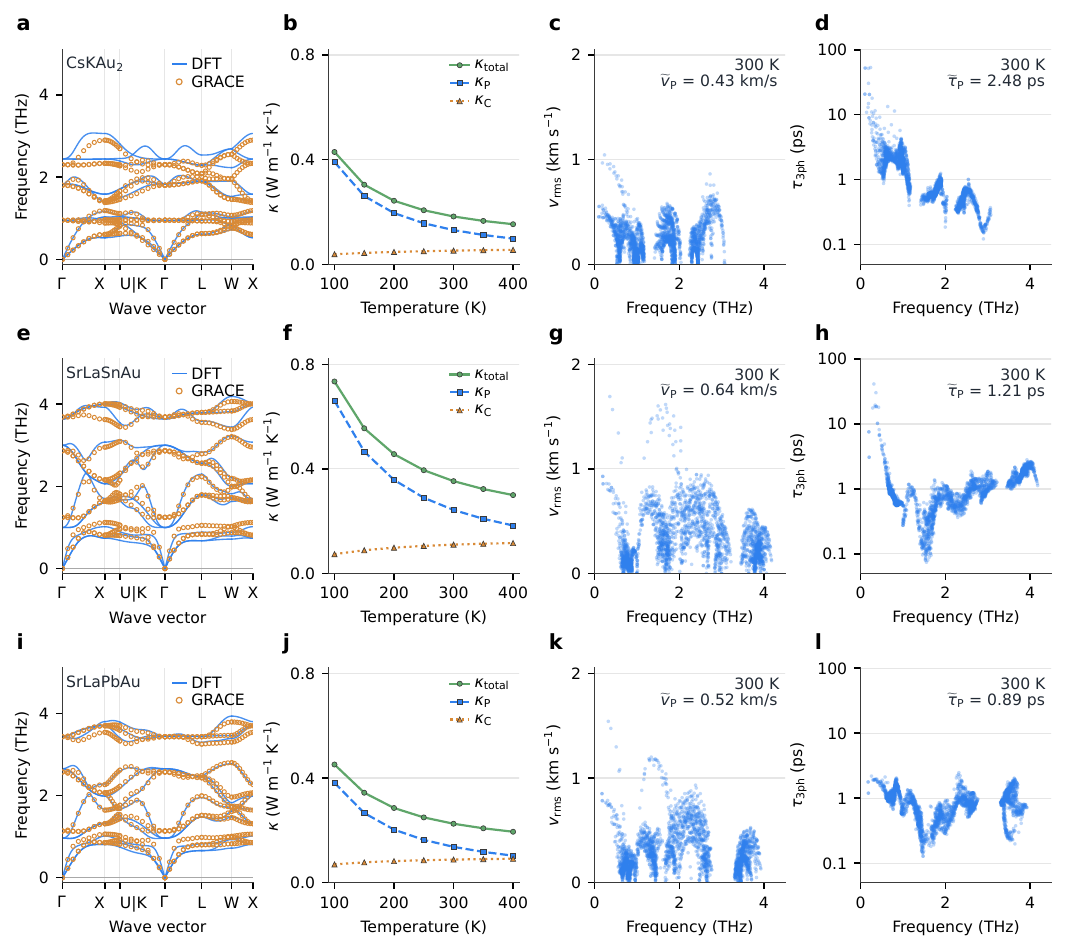}
\caption{\textbf{Phonon and lattice-thermal-transport properties of representative ultralow-conductivity crystals.}
Rows show \chem{CsKAu_2} (\textbf{a--d}), \chem{SrLaSnAu} (\textbf{e--h}), and \chem{SrLaPbAu} (\textbf{i--l}).
\textbf{a,e,i}, DFT phonon dispersions (blue lines) and GRACE predictions (orange open symbols).
\textbf{b,f,j}, Temperature-dependent total (green), population (blue), and coherence (orange) thermal conductivities.
\textbf{c,g,k}, DFT phonon group velocities as a function of frequency.
\textbf{d,h,l}, Three-phonon lifetimes at 300~K as a function of frequency.}
\label{fig:representative-transport}
\end{fixedfigure}

\begin{fixedfigure}
\centering
\includegraphics[width=\textwidth,height=.69\textheight,keepaspectratio]{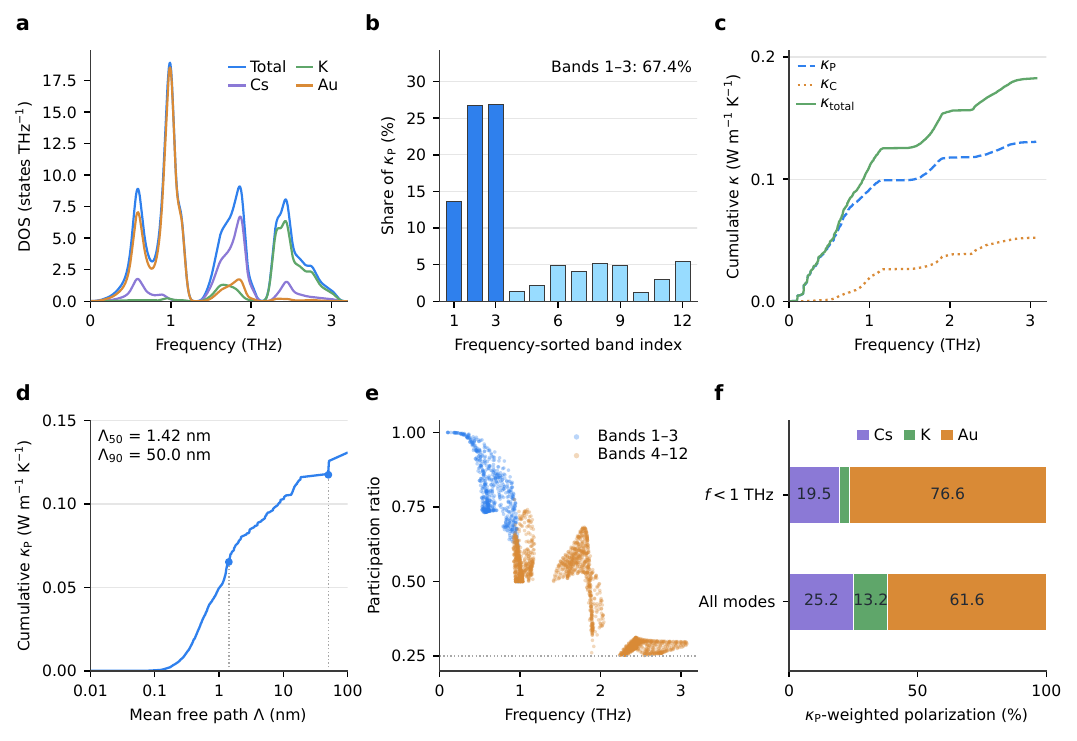}
\caption{\textbf{Phonon characteristics and heat-transport mechanisms in \chem{CsKAu_2}.}
\textbf{a}, Element-projected phonon density of states.
\textbf{b}, Branch-resolved particle thermal conductivity.
\textbf{c}, Cumulative particle (blue) and coherence (orange) thermal conductivities as a function of frequency.
\textbf{d}, Cumulative particle thermal conductivity as a function of phonon mean free path.
\textbf{e}, Displacement participation ratios of all phonon modes and modes below 1~THz.
\textbf{f}, Particle-thermal-conductivity-weighted atomic eigenvector projections for Cs, K, and Au.}
\label{fig:CsKAu2}
\end{fixedfigure}

To examine the phonon characteristics and microscopic origins of ultralow thermal conductivity, we first analyze the phonon dispersions of \(\mathrm{CsKAu_2}\), \(\mathrm{SrLaSnAu}\), and \(\mathrm{SrLaPbAu}\). As shown in Fig.~\ref{fig:representative-transport}a, e, and i, none of the three compounds exhibits appreciable imaginary phonon modes in the DFT phonon spectra, indicating dynamical stability within the harmonic approximation. GRACE reproduces the main features of the DFT phonon dispersions, while yielding lattice thermal conductivities of \(0.1752\), \(0.4636\), and \(0.3983~\mathrm{W\,m^{-1}\,K^{-1}}\), compared with the corresponding DFT values of \(0.1826\), \(0.3530\), and \(0.2254~\mathrm{W\,m^{-1}\,K^{-1}}\) (Table~\ref{tab:dft-low-kappa}). These comparisons support the use of GRACE for evaluating the phonon transport properties of these compounds. \(\mathrm{CsKAu_2}\) crystallizes in the \(Fm\overline{3}m\) space group (No. 225), whereas \(\mathrm{SrLaSnAu}\) and \(\mathrm{SrLaPbAu}\) both belong to \(F\overline{4}3m\) (No. 216).

Among the 58 DFT-validated compounds, \(\mathrm{CsKAu_2}\) exhibits the lowest lattice thermal conductivity, with a calculated value of \(0.1826~\mathrm{W\,m^{-1}\,K^{-1}}\) at 300 K. Its particle-transport-weighted median root-mean-square phonon group velocity is \(0.425~\mathrm{km\,s^{-1}}\) (Fig.~\(\ref{fig:representative-transport}\)b–d). Frequency-resolved analysis shows that modes below 1 THz contribute approximately \(70.0\%\) of the particle thermal conductivity. Further mode-resolved analysis reveals pronounced Au vibrational character in these low-frequency modes (Fig.~\(\ref{fig:CsKAu2}\)a,c,f), along with largely collective atomic displacements within the primitive cell (Fig.~\(\ref{fig:CsKAu2}\)e). Although low-frequency modes exhibit relatively low typical group velocities, their three-phonon lifetimes are longer than those of mid- and high-frequency modes, indicating that their transport characteristics cannot be attributed simply to enhanced three-phonon scattering. Detailed frequency-resolved mode statistics and comparisons of the two
weighting schemes are provided in Appendix~\ref{sec:si-frequency-mode-statistics},
Figs.~\ref{fig:si-frequency-v-tau-C-kappa} and~\ref{fig:si-velocity-distribution},
and Tables~\ref{tab:si-frequency-window} and~\ref{tab:si-kappa-weighted}. In addition, coherence transport accounts for approximately \(28.5\%\) of the total lattice thermal conductivity (Fig.~\(\ref{fig:CsKAu2}\)c), highlighting the importance of both particle and coherence contributions to heat transport in this material.

The isostructural compounds \(\mathrm{SrLaSnAu}\) and \(\mathrm{SrLaPbAu}\) provide a useful comparison for examining the influence of elemental substitution. They share the same stoichiometry and \(F\overline{4}3m\) space group, differing only by Sn/Pb substitution at the main-group element site. Their phonon group velocities and three-phonon lifetimes exhibit similar overall frequency-dependent distributions (Fig.~\ref{fig:representative-transport}e–l). Both compounds show low group velocities and finite phonon lifetimes, which jointly constrain phonon heat transport. Upon replacing Sn with Pb, the lattice thermal conductivity decreases from \(0.3530\) to \(0.2254~\mathrm{W\,m^{-1}\,K^{-1}}\), a reduction of 36.2\%. This decrease is accompanied by a reduction in the particle-transport-weighted median group velocity from \(0.643\) to \(0.515~\mathrm{km\,s^{-1}}\) and in the median three-phonon lifetime from \(1.210\) to \(0.893~\mathrm{ps}\). Correspondingly, the particle contribution decreases from \(0.2428\) to \(0.1370~\mathrm{W\,m^{-1}\,K^{-1}}\), accounting for approximately 82.9\% of the reduction in total thermal conductivity. The coherence contribution also decreases, from \(0.1102\) to \(0.0884~\mathrm{W\,m^{-1}\,K^{-1}}\), although its relative share increases from 31.2\% to 39.2\% because the particle contribution decreases more substantially. These results show that Pb substitution is accompanied by slower phonon propagation and shorter three-phonon lifetimes, with the reduction in particle transport accounting for most of the decrease in total thermal conductivity.

Despite their different group-velocity and lifetime characteristics, all three compounds exhibit both particle and coherence contributions to heat transport. At 300 K, particle transport remains dominant, while coherence transport accounts for 28.5\%–39.2\% of the total thermal conductivity. Between 100 and 400 K, the total lattice thermal conductivity decreases with increasing temperature in all three compounds, as the reduction in particle transport outweighs the increase in coherence transport. Overall, these results highlight low phonon group velocities as a common feature of the three ultralow-thermal-conductivity compounds, while the relative roles of phonon lifetimes and coherence transport depend on the specific material.

\subsection{Low lattice conductivity and crystalline-topological electronic states in cubic BaPd}\label{sec:bapd-topology}

\begin{fixedfigure}
\centering
\includegraphics[width=\textwidth,height=.69\textheight,keepaspectratio]{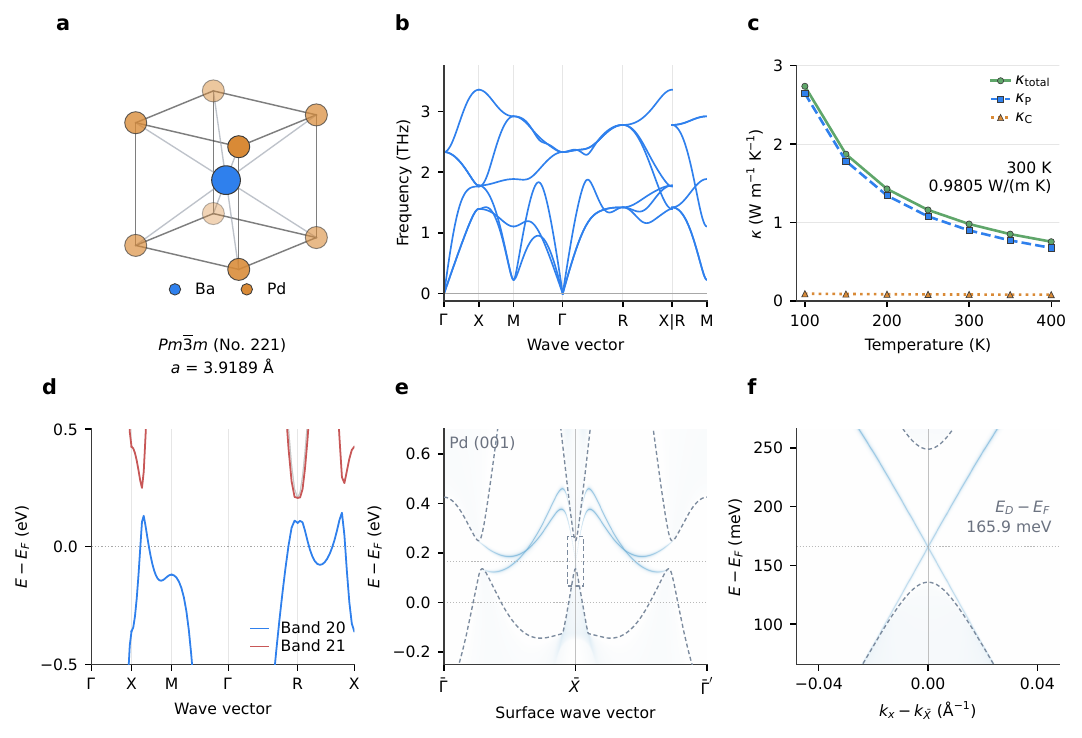}
\caption{\textbf{Low lattice thermal conductivity and SOC electronic surface states of cubic BaPd.}
\textbf{a}, Crystal structure of BaPd in the $Pm\overline{3}m$ space group.
\textbf{b}, Harmonic phonon dispersion.
\textbf{c}, Temperature-dependent total lattice thermal conductivity and its population and coherence contributions.
\textbf{d}, Bulk electronic band structure calculated with PBE+SOC.
\textbf{e}, Surface spectral function of the Pd-terminated (001) surface.
\textbf{f}, Enlarged view of the surface spectrum near $\overline{\mathrm X}$, showing a Dirac-type crossing.}
\label{fig:bapd-thermal-electronic}
\end{fixedfigure}

The cubic compound BaPd crystallizes in the $Pm\bar{3}m$ space group (No.~221) with two atoms in the primitive cell and a relaxed lattice parameter of approximately 3.9189~\AA\ (Fig.~\ref{fig:bapd-thermal-electronic}a). This minimal cell and high cubic symmetry make BaPd a structurally simple platform for phonon transport and electronic topology. The harmonic phonon dispersion shows no imaginary frequencies, confirming dynamical stability (Fig.~\ref{fig:bapd-thermal-electronic}b). The low-lying optical branches overlap with the acoustic branches below approximately 2~THz, consistent with the relatively small mass contrast between Ba and Pd. This overlap may open additional three-phonon scattering channels and shorten phonon lifetimes, contributing to the low thermal conductivity. At 300~K, the total lattice thermal conductivity is $\kappa_{\mathrm{L}} = 0.9805~\mathrm{W\,m^{-1}\,K^{-1}}$, with $\kappa_{\mathrm{P}} = 0.8994~\mathrm{W\,m^{-1}\,K^{-1}}$ (91.7\%) and $\kappa_{\mathrm{C}} = 0.0811~\mathrm{W\,m^{-1}\,K^{-1}}$ (8.3\%) (Fig.~\ref{fig:bapd-thermal-electronic}c). From 100 to 400~K, $\kappa_{\mathrm{L}}$ decreases monotonically from approximately 2.74 to 0.75~$\mathrm{W\,m^{-1}\,K^{-1}}$, following the $\sim 1/T$ trend characteristic of anharmonic scattering. This reduction is driven almost entirely by the population channel, while the coherence contribution remains nearly temperature-independent. In contrast to the four-atom compounds in Section~\ref{sec:representative-transport}, whose coherence fractions range from 28.5\% to 39.2\%, BaPd is population-dominated.

Beyond its thermal transport properties, BaPd exhibits nontrivial crystalline topology when spin--orbit coupling (SOC) is included. The PBE+SOC band structure (Fig.~\ref{fig:bapd-thermal-electronic}d) reveals a narrow indirect gap of approximately $62.7~\mathrm{meV}$ between bands 20 and 21, opened by the strong SOC of the heavy Ba and Pd atoms. Topological analysis yields mirror Chern numbers $C_m = 2$ on all three mirror-symmetric planes, classifying BaPd as a topological crystalline insulator (TCI). This nontrivial topology is directly manifested at the surface. The spectral function of the Pd-terminated (001) surface (Fig.~\ref{fig:bapd-thermal-electronic}e) shows topologically protected surface states that bridge the projected bulk gap, connecting the valence and conduction continua---a hallmark of bulk--boundary correspondence. An enlarged view near the surface $\bar{X}$ point (Fig.~\ref{fig:bapd-thermal-electronic}f) resolves a clear Dirac-type crossing at approximately $165.9~\mathrm{meV}$ above the Fermi energy. The position of this Dirac point within the projected gap suggests tunability via charge doping or gating. Taken together, these results establish BaPd as a rare example of a simple cubic compound that simultaneously hosts ultralow lattice thermal conductivity and topologically protected electronic surface states. The coexistence of strong phonon scattering and symmetry-protected Dirac surface states in a single two-atom crystal makes BaPd a compelling candidate for thermoelectric applications where topological surface channels may complement the suppressed lattice heat transport.

\subsection{Thermal-conductivity prediction models and benchmark evaluation}\label{sec:benchmark}

\begin{fixedfigure}
\includegraphics[
  width=.85\textwidth,
  height=.50\textheight,
  keepaspectratio
]{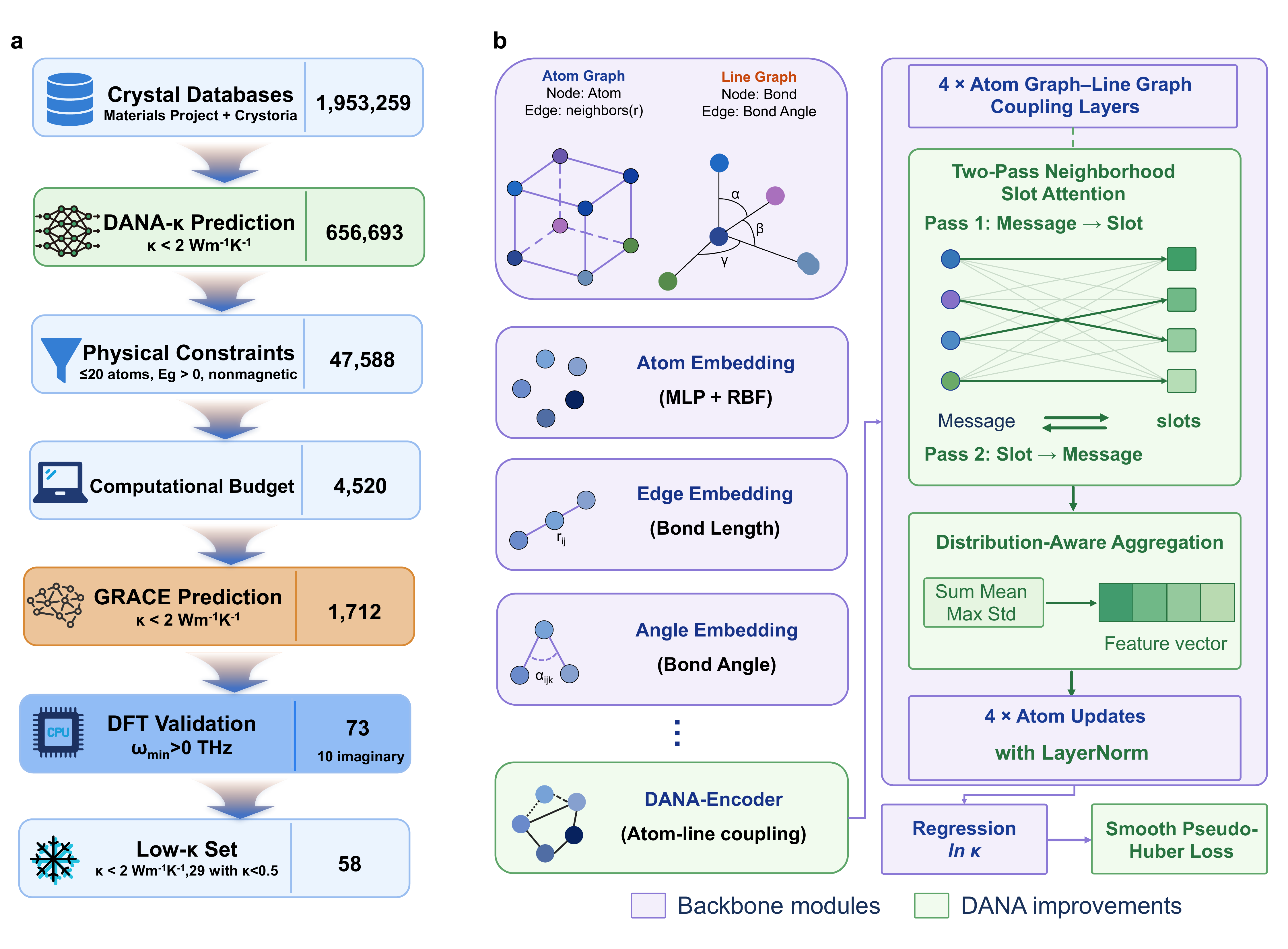}
\caption{\textbf{Multilevel screening workflow and DANA-$\kappa$ architecture.}
\textbf{a}, High-throughput screening workflow. DANA-$\kappa$ predicts the thermal conductivity of 1,953,259 crystal structures, followed by physical-constraint and computational-budget filtering, retaining 4,520 candidates. Further screening with GRACE and Phono3py identifies 1,712 low-thermal-conductivity candidates. DFT validation of 73 candidates ultimately yields 58 structures with lattice thermal conductivities below $2\ \mathrm{W\,m^{-1}\,K^{-1}}$, including 29 below $0.5\ \mathrm{W\,m^{-1}\,K^{-1}}$.
\textbf{b}, DANA-$\kappa$ architecture. Atomic graph and line graphs are coupled through two-pass neighborhood slot attention and distribution-aware aggregation to extract structural features, followed by atom-update layers and a regression head to predict $\ln\kappa_{\mathrm L}$. The model is trained using a smooth pseudo-Huber loss. Purple and green denote backbone modules and DANA-specific improvements, respectively.}
\label{fig:architecture-screening}
\end{fixedfigure}

\begin{fixedtable}
\centering
\small
\caption{Benchmark of thermal-conductivity surrogate models. Errors are reported as the mean absolute error (MAE) of the natural logarithm of thermal conductivity and are dimensionless. DANA-$\kappa$ results are from this work; the remaining benchmark results are reproduced from Wang et al.~\cite{Surrogate2026}}
\label{tab:model-benchmark}
\begin{tabular}{@{}lrrrr@{}}
\toprule
Model & Random & Space group & OOD & Avg \\
\midrule
DANA-$\kappa$ & \textbf{0.370} & \textbf{0.480} & \textbf{1.182} & \textbf{0.677} \\
ALiEGNN & 0.379 & 0.512 & 1.245 & 0.712 \\
Orb + CNN & 0.440 & 0.510 & 1.404 & 0.784 \\
HackNIP & 0.380 & 0.501 & 1.516 & 0.799 \\
ViKING & 0.487 & 0.597 & 1.370 & 0.818 \\
CGCNN & 0.523 & 0.652 & 1.294 & 0.823 \\
KAN + Physical-informed feat. & 0.473 & 0.596 & 1.424 & 0.831 \\
MLP + Physical-informed feat. & 0.431 & 0.627 & 1.454 & 0.837 \\
MACE(OMAT)+MLP & 0.448 & 0.601 & 1.466 & 0.838 \\
MACE(MPA)+MLP & 0.445 & 0.602 & 1.484 & 0.843 \\
WyFormer & 0.503 & 0.625 & 1.419 & 0.849 \\
XGBoost + Physical-informed feat. & 0.444 & 0.540 & 1.610 & 0.865 \\
Crabnet & 0.523 & 0.674 & 1.422 & 0.873 \\
Orb + TabPFN & 0.378 & 0.494 & 1.764 & 0.879 \\
WyCryst+ & 0.491 & 0.700 & 1.516 & 0.902 \\
\bottomrule
\end{tabular}
\end{fixedtable}

\begin{fixedtable}
\centering
\small
\setlength{\tabcolsep}{10pt}
\renewcommand{\arraystretch}{1.08}
\caption{On the 103-crystal thermal-conductivity benchmark reported by P\'ota et al.~\cite{Pota2026}, we additionally evaluate GRACE-2L-OAM and MatterSim-V1-5M and include their results as two additional rows. GRACE-2L-OAM achieves the best predictive performance on this benchmark. SRE denotes the symmetric relative error in total thermal conductivity, defined as $2|\kappa_{\mathrm{MLIP}}-\kappa_{\mathrm{DFT}}|/(\kappa_{\mathrm{MLIP}}+\kappa_{\mathrm{DFT}})$. SRME instead quantifies the normalized sum of absolute differences between mode-resolved thermal-conductivity contributions, thereby capturing compensating errors. Both metrics are dimensionless.}
\label{tab:foundation-potential-benchmark}
\begin{tabular}{@{}lrrr@{}}
\toprule
Potential version & Valid / 103 & Mean SRME $\downarrow$ & Mean SRE $\downarrow$ \\
\midrule
\textbf{GRACE-2L-OAM} & \textbf{103/103} & \textbf{0.294} & \textbf{0.144} \\
MACE-MPA-0 & 103/103 & 0.412 & 0.205 \\
GRACE-1L-OAM & 103/103 & 0.517 & 0.282 \\
GRACE-2L-r6-11Nov2024 & 100/103 & 0.525 & 0.319 \\
MatterSim-V1-1M & 102/103 & 0.555 & 0.343 \\
MACE-agnesi-medium & 100/103 & 0.565 & 0.378 \\
MatterSim-V1-5M & 100/103 & 0.575 & 0.413 \\
MACE-L2 & 99/103 & 0.670 & 0.528 \\
MACE-M & 100/103 & 0.682 & 0.471 \\
SevenNet & 95/103 & 0.762 & 0.605 \\
MACE-L & 97/103 & 0.932 & 0.719 \\
M3GNet & 96/103 & 1.409 & 1.340 \\
CHGNet & 87/103 & 1.717 & 1.696 \\
ORBv2-MPTraj & 100/103 & 1.726 & 1.454 \\
ORBv2 & 103/103 & 1.734 & 1.488 \\
eqV2-full-S & 97/103 & 1.771 & 1.483 \\
ORBv1-MPTraj & 93/103 & 1.787 & 1.555 \\
\bottomrule
\end{tabular}
\par\vspace{6pt}
\end{fixedtable}

We use DANA-$\kappa$ for large-scale screening and GRACE-2L-OAM for subsequent phonon-transport calculations. DANA-$\kappa$ directly predicts lattice thermal conductivity from crystal structures, without explicitly calculating force constants or solving the phonon Boltzmann transport equation, making it suitable for screening large candidate pools. In contrast, GRACE-2L-OAM provides an interatomic potential for accelerating atomistic calculations and is used in combination with phonon-transport methods to evaluate thermal conductivity.

DANA-$\kappa$ is built on an atom-graph and bond-angle line-graph architecture, with two-pass neighborhood slot attention and distribution-aware aggregation. The former captures interactions among neighboring atomic environments, while the latter summarizes the distribution of messages using multiple statistical descriptors. The model is trained on 6,966 inorganic crystals from the Phonix database, using their first-principles lattice thermal conductivities at 300 K as targets. We evaluate its performance on random, space-group-disjoint, and low-conductivity out-of-distribution (OOD) test sets, which probe interpolation, structural generalization, and extrapolation to low-conductivity materials, respectively.

As shown in Table~\ref{tab:model-benchmark}, DANA-$\kappa$ achieves mean absolute errors (MAEs) in $\ln\kappa_{\mathrm L}$ of 0.369, 0.479, and 1.181 on the three test sets, respectively, with an average MAE of 0.677. This is lower than that of the ALiEGNN baseline (0.711), supporting its use for large-scale screening. The higher error on the low-conductivity OOD set also indicates the difficulty of extrapolating to this regime. We further assess the interatomic potentials used in phonon-transport calculations on the 103-crystal benchmark reported by P\'ota et al.~\cite{Pota2026}. We add GRACE-2L-OAM and MatterSim-V1-5M to the original benchmark. As reported in Table~\ref{tab:foundation-potential-benchmark}, GRACE-2L-OAM produces valid results for all 103 crystals and achieves the lowest mean SRME (0.294) and SRE (0.144) among the evaluated models. MatterSim-V1-5M gives valid results for 100 crystals, with mean SRME and SRE values of 0.575 and 0.413, respectively. These results motivate our selection of GRACE-2L-OAM for the subsequent phonon-transport calculations. The two benchmarks assess different aspects of the workflow: the first evaluates direct structure-to-thermal-conductivity prediction, whereas the second evaluates the performance of interatomic potentials in phonon-transport calculations. Their error metrics are therefore not directly comparable.

Although DANA-$\kappa$ substantially improves large-scale screening efficiency, it systematically overestimates the thermal conductivity of ultralow-$\kappa$ materials. Consequently, applying a strict threshold of $\hat{\kappa}_L<2~\mathrm{W\,m^{-1}\,K^{-1}}$ alone may miss some genuinely low-$\kappa$ candidates. To quantify the extent of such false negatives, we divided the screening into a priority tier ($\hat{\kappa}_L<2$) and an extended tier ($2\leq\hat{\kappa}_L<8$), and retrospectively evaluated both tiers using existing GRACE transport and DFT results (Appendix~\ref{app:computational-budget}
and Supplementary Fig.~\ref{fig:Fig-screening-two_thresholds-2-and-8}). Of the 59 candidates completed under the same DFT workflow, 58 ultimately exhibited $\kappa_{\rm DFT}<2~\mathrm{W\,m^{-1}\,K^{-1}}$, including 50 from the priority tier and 8 from the extended tier. These results demonstrate that the extended threshold primarily recovers candidates missed due to model overestimation, rather than serving as an independent criterion for identifying low thermal conductivity. We note that the DANA predictions reported in Table~\ref{tab:dft-low-kappa} were obtained using DFT-relaxed structures, whereas the screening tiers were defined based on the initial database structures; thus, the two involve different structural configurations. Of the nine materials with DANA predictions above
$2~\mathrm{W\,m^{-1}\,K^{-1}}$ in Table~\ref{tab:dft-low-kappa},
seven belonged to the extended tier on their initial database structures,
whereas BaBrCl and RbBaAs belonged to the priority tier and their predictions
increased after relaxation. Conversely, RbBaBi belonged to the extended
tier but its predicted conductivity decreased to
$1.9531~\mathrm{W\,m^{-1}\,K^{-1}}$ after DFT relaxation.

\subsection{Discussion}

In summary, we establish a multilevel high-throughput screening framework that combines DANA-$\kappa$, machine-learning interatomic potentials, and DFT calculations to efficiently identify crystals with ultralow lattice thermal conductivity from large-scale structural databases. This approach substantially reduces the computational cost of screening and enables the identification of 58 previously unreported low-thermal-conductivity crystals. Nevertheless, the present search is primarily restricted to structures available in existing databases. Recent advances in generative models for crystal structures, represented by MatterGen~\cite{MatterGen2025}, DiffCSP~\cite{DiffCSP2023}, and CrystalFormer~\cite{CrystalFormer2025}, have opened new opportunities for AI-driven inverse materials design~\cite{AIInverseDesignReview2025,NatureMaterialsInverseDesign2026,PhononBench2025} by enabling the exploration of chemical and structural spaces beyond known materials. Frameworks such as InvDesFlow~\cite{InvDesFlow2025} and InvDesFlow-AL~\cite{InvDesFlowAL2025} have demonstrated the potential of this approach in superconducting materials discovery, identifying numerous promising candidates, including $\mathrm{Li_2AuH_6}$ with a predicted superconducting transition temperature of approximately 140 K at ambient pressure~\cite{Li2AuH6NC2025,Li2AuH6PRB2025}. More recently, generative and AI-assisted approaches have begun to address thermal transport properties directly, enabling the discovery of crystals with ultralow lattice thermal conductivity through generative crystal design and cross-property transfer learning~\cite{Minhas2026}, diffusion-based inverse design of thermoelectric materials~\cite{WangTEID2026}, and LLM-assisted exploration of ultralow-$\kappa$ chalcogenides~\cite{LiuUltralow2026}. Generative learning has also been applied to the discovery of materials with ultrahigh lattice thermal conductivity~\cite{GuoGenerative2025}. These developments highlight the complementary roles of database-driven screening and generative inverse design. Integrating crystal generation with the present multilevel screening framework could enable a more comprehensive exploration of materials beyond existing databases and accelerate the discovery of crystals with targeted thermal transport properties.

Scientific agents offer a complementary route to automating materials-discovery workflows. General-purpose coding agents, such as OpenAI Codex and Claude Code, together with scientific agent frameworks, including Coscientist~\cite{Coscientist}, ChemCrow~\cite{ChemCrow}, SciAgents~\cite{SciAgents}, and AtomAgents~\cite{AtomAgents}, have demonstrated the potential of LLM-based agents for scientific reasoning, tool integration, and collaborative problem-solving. In computational materials science, VASPilot~\cite{VASPilot} automates end-to-end first-principles calculation workflows, while PhysAgent~\cite{PhysAgent} extends multi-agent collaboration to physics modeling, DFT simulations, and scientific data analysis. By integrating these agents with crystal generative models, machine-learning interatomic potentials, and first-principles simulation tools, it may become possible to establish autonomous workflows that connect candidate generation, structural optimization, property evaluation, and iterative feedback. This integration could reduce manual intervention in iterative materials discovery while retaining first-principles validation of candidate properties.

\section{Method}\label{sec:method}

DANA-$\kappa$ is built on the atom-graph and line-graph architecture of ALiEGNN~\cite{Surrogate2026,ALIGNN2021}. A crystal is represented as a directed multigraph $G=(V,E)$, where atoms are nodes and edges $e:j\to i$ connect periodic neighboring atoms. The corresponding line graph $L(G)$ represents bonds as nodes and bond angles as edges. The network consists of eight atom-update layers. The first four layers couple atom-graph and line-graph updates, whereas the remaining four perform atom-graph updates without further line-graph updates.

At each atom-update layer, edge messages are constructed from the representations of the source and target atoms, together with geometric and bond features:
\begin{equation}
\mathbf m_e =
\phi_{\mathrm e}
\left(
\mathbf h_j \Vert \mathbf h_i
\Vert \mathbf s_e \Vert \mathbf b_e
\right),
\qquad e:j\to i,
\label{eq:edge-message}
\end{equation}
where $\mathbf h_i$ denotes the atomic representation, $\mathbf s_e$ contains geometric features, and $\mathbf b_e$ represents bond features updated through the line graph. Layer indices are suppressed for clarity.

DANA-$\kappa$ introduces two architectural improvements to this backbone. First, two-pass neighborhood slot attention enables messages to incorporate contextual information from the local atomic environment. For each central atom, learnable slots first aggregate its incoming edge messages into a set of neighborhood representations. Each edge message then retrieves context from these slots, and the retrieved context is used to refine the message through a gated residual connection:
\begin{equation}
\begin{aligned}
\mathbf z_{is}
&= \sum_{e\in\mathcal E_i} a_{es}\mathbf v_e,\\
\mathbf c_e
&= \sum_{s=1}^{K} b_{es}\mathbf z_{is},\\
\widetilde{\mathbf m}_e
&= \mathbf m_e+
\sigma(\alpha)W_{\mathrm O}
\operatorname{SiLU}(\mathbf c_e-\mathbf v_e),
\end{aligned}
\label{eq:slot-attention}
\end{equation}
where $\mathcal E_i$ denotes the incoming edges of atom $i$, $K$ is the number of slots, and $\mathbf v_e=W_{\mathrm V}\mathbf m_e$ is the projected value vector. The vector $\mathbf z_{is}$ summarizes the neighborhood information captured by slot $s$, while $\mathbf c_e$ is the context retrieved by edge $e$. The attention weights $a_{es}$ and $b_{es}$ are softmax-normalized over incoming edges and slots, respectively. Here, $W_{\mathrm O}$ maps the retrieved context to the message dimension, and $\alpha$ is a learnable scalar for each layer. The refined messages are used in node aggregation, while the original messages are retained for the inherited bond-residual and internal-coordinate-feature updates. This mechanism captures neighborhood-level interactions without constructing a dense pairwise attention matrix.

Second, distribution-aware multi-statistic aggregation replaces conventional sum aggregation with a combination of four complementary statistics of the refined messages:

\begin{align}
\mathbf S_i &=
\sum_{e\in\mathcal E_i}\widetilde{\mathbf m}_e,\\
\boldsymbol\mu_i &=
\frac{1}{n_i}\sum_{e\in\mathcal E_i}\widetilde{\mathbf m}_e,\\
\mathbf M_i &=
\max_{e\in\mathcal E_i}\widetilde{\mathbf m}_e,\\
\mathbf D_i &=
\operatorname{Std}_{\mathrm{smooth,pop}}
\left(
\{\widetilde{\mathbf m}_e:e\in\mathcal E_i\}
\right),
\label{eq:distribution-aggregation}
\end{align}

where $n_i=|\mathcal E_i|$, and all statistics are computed element-wise. The four terms describe the channel-wise total, mean, maximum, and variation of neighboring messages, respectively. The standard deviation is evaluated using a smoothed population-variance estimator. These statistics are concatenated with the central atom representation and passed to the node-update network:
\begin{equation}
\mathbf h_i' =
\phi_{\mathrm n}
\left(
\mathbf h_i\Vert
\mathbf S_i\Vert
\boldsymbol\mu_i\Vert
\mathbf M_i\Vert
\mathbf D_i
\right).
\label{eq:distribution-node-update}
\end{equation}
For atoms without incoming edges, all four statistics are set to zero. By combining these statistics, the model represents not only the aggregate magnitude of local messages but also their average response, extreme values, and channel-wise variation.

Finally, mean pooling over the updated atomic representations produces a crystal-level representation, which is mapped to the logarithmic lattice thermal conductivity:
\begin{equation}
\widehat y =
\mathbf w^{\mathsf T}
\left(
\frac{1}{|V|}\sum_{i\in V}\mathbf h_i^{(8)}
\right)+b.
\label{eq:crystal-readout}
\end{equation}
The target is defined as
$y=\ln(\kappa_{\mathrm L}/\kappa_0)$, where
$\kappa_0=1~\mathrm{W\,m^{-1}\,K^{-1}}$.
The predicted conductivity is recovered as
$\widehat{\kappa}_{\mathrm L}=\kappa_0\exp(\widehat y)$.
The model is trained by minimizing a smooth pseudo-Huber loss on the logarithmic conductivity target.

First-principles validation of the selected low-conductivity structures was
performed using VASP 6.5.0~\cite{VASP1996,Kresse1996} with the
projector augmented-wave method and the PBE
functional~\cite{PBE1996} in a non-spin-polarized approximation.
The plane-wave cutoff was set to $1.3$ times the largest ENMAX of the
selected PAW datasets for each material, spanning 286.15--525.11~eV for
the 58 low-conductivity structures. Primitive cells initialized from the
corresponding GRACE-relaxed geometries were optimized at zero external
pressure while retaining the target space group, using an electronic
convergence threshold of $10^{-8}$~eV per cell and a target force threshold
of $10^{-4}$~eV/\AA{}, followed by independent static checks.
Primitive-cell relaxation/static calculations and displaced-supercell
force calculations for second- and third-order force constants used
$\Gamma$-centered electronic meshes of $6\times6\times6$,
$2\times2\times2$, and $3\times3\times3$, respectively.
Force constants were obtained by finite displacements of 0.03~\AA{}
in $4\times4\times4$ and $2\times2\times2$ primitive-cell supercells.
Thermal conductivities were evaluated at 300~K using phono3py
2.3.2~\cite{Phono3py2023,Phono3py2015} on a
$19\times19\times19$ phonon mesh, with tetrahedron integration and the
Wigner formulation using a relaxation-time treatment of the population
term~\cite{Simoncelli2019,Simoncelli2022}. Isotope scattering and a 1~m
boundary mean free path were included; spin--orbit coupling and
non-analytical corrections were omitted from the lattice-transport
calculations. The reported scalar is the diagonal average of the total
population-plus-coherence conductivity tensor, excluding electronic
conductivity. Shared calculation settings and material-specific PAW datasets, cutoffs,
and displacement counts are provided in Appendix~\ref{app:dft-methods}
and Tables~\ref{tab:sm-common-dft}--\ref{tab:sm-dft-parameters-2}.

Before submitting the final DFT calculations, we conducted a novelty assessment of each candidate through database deduplication and individual literature searches. This assessment confirmed that, within the scope of our search, no lattice thermal conductivity calculations had been reported for the target crystal phases of the 58 newly identified structures. Database screening used the 103 benchmark crystals reported by P\'ota et al.~\cite{Pota2026} (reduced to 70 unique chemical formulas) as references. After standardizing database identifiers and aliases and reducing chemical formulas, candidate and reference structures of identical composition were compared using \texttt{pymatgen}'s \texttt{StructureMatcher}, allowing primitive-cell reduction and volume scaling but excluding elemental substitutions, subset matching, and supercell searches. Before each DFT submission batch, we also searched publicly available literature using chemical formulas, element orderings, and compound names, combined with keywords including \textit{thermal conductivity}, \textit{lattice thermal conductivity}, \textit{phonon transport}, and \textit{Boltzmann transport}. Studies of compounds with matching compositions were further examined to determine whether they reported lattice thermal conductivity calculations for the target crystal phases.

High-throughput potential-based screening employed the pretrained
GRACE-2L-OAM model~\cite{GRACE2026}, specifically the
\texttt{GRACE\_2L\_OAM\_28Jan25} checkpoint accessed through
TensorPotential 0.4.5. Standardized conventional cells were relaxed using
a two-stage FIRE optimization with the ASE
\texttt{FrechetCellFilter}; both stages used
$f_{\max}=10^{-4}$~eV/\AA{} and a maximum of 300 steps, with symmetry
monitoring and retention of the symmetry-preserving structure when
required. Forces for the displaced supercells were evaluated directly
with GRACE, and force constants and conductivity were obtained using
phono3py 3.24.0. The second- and third-order supercells were
$4\times4\times4$ and $2\times2\times2$ multiples of the conventional
cell, respectively, whereas the DFT supercells were constructed from
the primitive cell. The potential and DFT calculations used their
respective relaxed geometries, with the same displacement amplitude,
phonon mesh, temperature, Wigner/RTA treatment, isotope scattering,
and boundary mean free path. These uniform settings define a screening
protocol; convergence checks for individual materials remain necessary.

\section{Code and data availability}
The DANA-$\kappa$ model, source code, and datasets used in this work are publicly available at GitHub: \url{https://github.com/xqh19970407/DaNa-kappa}.

\noindent\textbf{Acknowledgement:}
The work is supported by Beijing Natural Science Foundation(No.Z250005, No.Z260002), the National Natural Science Foundation of China (No.62476278, No.11934020), and the National Key R\&D Program of China (Grants No. 2024YFA1408601). Computational resources have been provided by the Physical Laboratory of High Performance Computing at Renmin University of China. 
\\

\noindent\textbf{Corresponding authors:} Correspondence and requests for materials should be addressed to Ze-Feng Gao (zfgao@ruc.edu.cn) and Zhong-Yi Lu (zlu@ruc.edu.cn). \\

\noindent\textbf{Competing interests:}
The authors declare no competing interests.\\

\noindent\textbf{Author's contributions:}
X.Q.H. and Z.F.G. conceived and designed the study and conducted the research. X.Q.H., Z.F.G., and Z.Y.L. drafted the original manuscript. All authors contributed to the research discussions.\\

\noindent\textbf{Supplementary materials:}
Supplementary material is provided in the Appendix.

\clearpage

\newpage

\onecolumn

\appendix

\setcounter{figure}{0}\renewcommand{\thefigure}{S\arabic{figure}}
\renewcommand{\theHfigure}{supp.\arabic{figure}}
\setcounter{table}{0}\renewcommand{\thetable}{S\arabic{table}}
\renewcommand{\theHtable}{supp.\arabic{table}}
\numberwithin{equation}{section}

{
\noindent\textbf{Contents of the Appendix}\par\medskip

\noindent
\hyperref[sec:si-frequency-mode-statistics]{\ref*{sec:si-frequency-mode-statistics}\quad Frequency-Resolved Phonon Mode Statistics of CsKAu$_2$}\dotfill\pageref{sec:si-frequency-mode-statistics}\par\smallskip

\noindent
\hyperref[app:dft-methods]{\ref*{app:dft-methods}\quad DFT and lattice-transport calculations}\dotfill\pageref{app:dft-methods}\par\smallskip

\noindent
\hyperref[app:computational-budget]{\ref*{app:computational-budget}\quad Empirical computational-budget filter}\dotfill\pageref{app:computational-budget}\par\smallskip

\noindent
\hyperref[app:Ablation]{\ref*{app:Ablation}\quad Ablation}\dotfill\pageref{app:Ablation}\par\smallskip

\noindent
\hyperref[app:all-transport]{\ref*{app:all-transport}\quad Phonon transport of the 58 low-conductivity phases}\dotfill\pageref{app:all-transport}\par
}
\clearpage

\section{Frequency-Resolved Phonon Mode Statistics of CsKAu$_2$}
\label{sec:si-frequency-mode-statistics}

\begin{table}[htbp]
\centering
\small
\caption{Mode-count fractions, heat-capacity fractions, and median values of four mode quantities in three frequency windows for $\mathrm{CsKAu_2}$ at 300~K. $v_{\rm rms}$, $\tau_3$, $C_\lambda/k_{\rm B}$, and single-mode $\kappa_{\rm P}$ are weighted by full-Brillouin-zone mode counts.}
\label{tab:si-frequency-window}
\resizebox{\textwidth}{!}{\begin{tabular}{lccccccc}
\hline
Window/THz & Mode fraction (\%) & $C_\lambda$ fraction (\%) & Median $v_{\rm rms}$ (km\,s$^{-1}$) & Median $\tau_3$ (ps) & Median $C_\lambda/k_{\rm B}$ & Median $\kappa_{\rm P}$ (W\,m$^{-1}$\,K$^{-1}$) & $\kappa_{\rm P}$ share (\%) \\
\hline
$<1$ & 34.63 & 34.79 & 0.1964 & 2.147 & 0.99865 & $1.029\times10^{-6}$ & 69.98 \\
$1$--$2$ & 39.84 & 39.88 & 0.2136 & 0.818 & 0.99441 & $4.445\times10^{-7}$ & 20.14 \\
$\ge 2$ & 25.52 & 25.33 & 0.2685 & 0.458 & 0.98725 & $4.684\times10^{-7}$ & 9.89 \\
\hline
\end{tabular}}
\end{table}

\begin{table}[htbp]
\centering
\small
\caption{$\kappa_{\rm P}$-weighted median group velocity and three-phonon lifetime. Here each mode is weighted by its particle thermal conductivity contribution $\kappa_{\rm P}$, so the statistics emphasize the modes that dominate heat transport. For comparison, Table~\ref{tab:si-frequency-window} reports mode-count-weighted medians. The low-frequency window shows higher $\kappa_{\rm P}$-weighted $v_{\rm rms}$ and $\tau_3$ than the other windows, indicating that the heat-carrying low-frequency modes are faster and longer-lived than the typical low-frequency mode.}
\label{tab:si-kappa-weighted}
\begin{tabular}{lcc}
\hline
Window/THz & $\kappa_{\rm P}$-weighted median $v_{\rm rms}$ (km\,s$^{-1}$) & $\kappa_{\rm P}$-weighted median $\tau_3$ (ps) \\
\hline
$<1$ & 0.4539 & 3.544 \\
$1$--$2$ & 0.3708 & 0.797 \\
$\ge 2$ & 0.4092 & 0.491 \\
\hline
\end{tabular}
\end{table}

\begin{fixedfigure}
\includegraphics[width=\textwidth,height=.85\textheight,keepaspectratio]{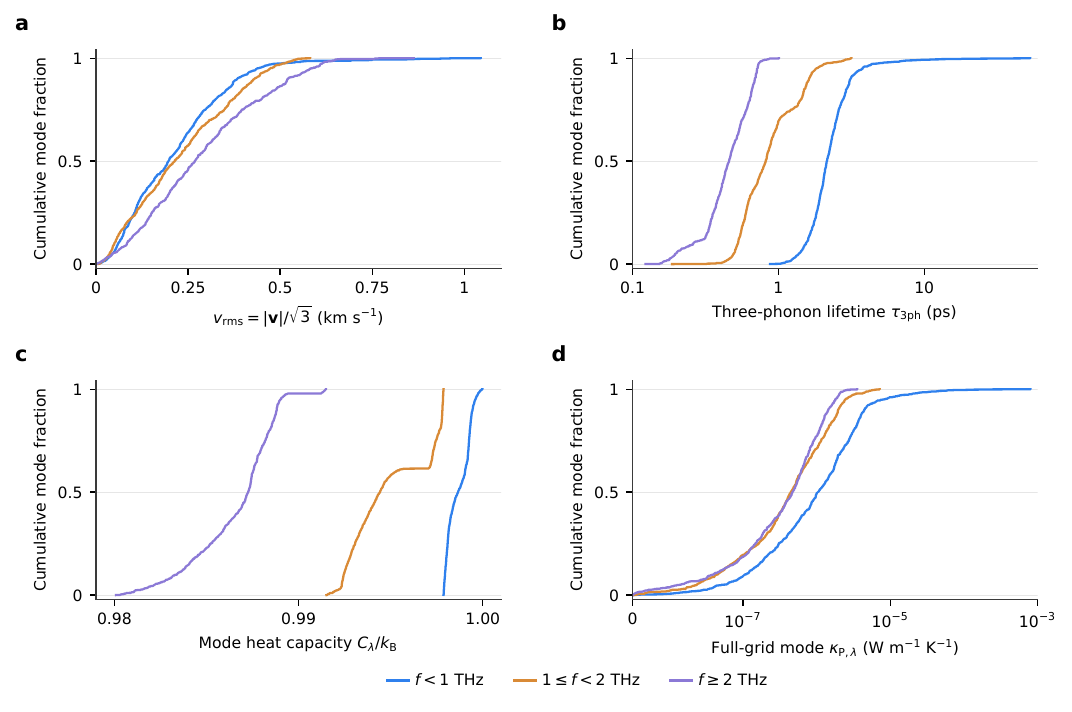}
\caption{Mode-resolved distributions of four phonon quantities for $\mathrm{CsKAu_2}$ at 300~K: (a) root-mean-square group velocity $v_{\rm rms}$; (b) three-phonon lifetime $\tau_3$; (c) mode heat capacity $C_\lambda/k_{\rm B}$; (d) particle thermal conductivity contribution $\kappa_{\rm P}$ per full-grid mode. Statistics are weighted by full-Brillouin-zone mode counts.}
\label{fig:si-frequency-v-tau-C-kappa}
\end{fixedfigure}

\begin{fixedfigure}
\includegraphics[width=\textwidth,height=.85\textheight,keepaspectratio]{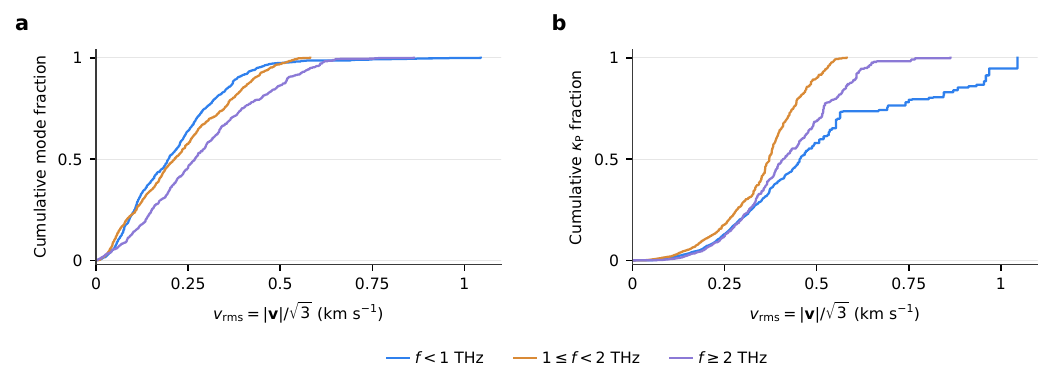}
\caption{Two weighting schemes for the group-velocity distribution of $\mathrm{CsKAu_2}$: (a) weighted by the number of modes represented by each $q$ star; (b) weighted by $\kappa_{\rm P}$. The different trends show that the typical velocity of all low-frequency modes and that of the low-frequency modes actually carrying heat are not interchangeable.}
\label{fig:si-velocity-distribution}
\end{fixedfigure}

Fig.~\ref{fig:si-frequency-v-tau-C-kappa} shows the frequency-window distributions of the four mode quantities, and Fig.~\ref{fig:si-velocity-distribution} compares the two weighting schemes for group velocity. With mode-count weighting, the median $v_{\rm rms}$ values for $<1$, $1$--$2$, and $\ge 2$~THz are 0.196, 0.214, and 0.268~km\,s$^{-1}$, respectively, while the corresponding median $\tau_3$ values are 2.147, 0.818, and 0.458~ps. Low-frequency modes thus have smaller typical velocities but longer lifetimes than mid- and high-frequency modes, indicating that their transport characteristics cannot be attributed simply to enhanced three-phonon scattering. The median mode heat capacities in all three windows remain close to $k_{\rm B}$, differing by only about 1\%. Modes below 1~THz account for only about 34.63\% of the mode count and 34.79\% of the heat capacity, yet contribute about 69.98\% of $\kappa_{\rm P}$, so their transport dominance cannot be explained by mode count or heat capacity alone.

With $\kappa_{\rm P}$ weighting, the low-frequency median $v_{\rm rms}$ and $\tau_3$ increase to 0.454~km\,s$^{-1}$ and 3.544~ps, higher than those in the other windows, indicating that the low-frequency modes actually carrying heat tend to be faster and longer-lived. Velocity and lifetime distributions are correlated across frequency windows, and products of medians do not constitute a strict decomposition of $\kappa$. This appendix discusses only the particle term $\kappa_{\rm P}$; the coherence term $\kappa_{\rm C}$ follows the main text.

\section{DFT and lattice-transport calculations}\label{app:dft-methods}
VASP 6.5.0/PAW--PBE calculations used non-spin-polarized states without SOC, with cutoffs of $1.3\max(\mathrm{ENMAX})$ for the selected PAW datasets~\cite{VASP1996,PBE1996}. Independent initial static calculations, segmented zero-external-pressure optimization and independent final self-consistent calculations with regenerated plane-wave bases were applied. Optimization targets and final structural convergence criteria are documented separately in Table~\ref{tab:sm-common-dft}; RbTlS has a final maximum force of $1.1217\times10^{-4}$ eV\,\AA$^{-1}$. Material-specific PAW datasets, plane-wave cutoffs and displacement counts are provided in the supplementary tables. 

Tables~\ref{tab:sm-dft-parameters-1} and~\ref{tab:sm-dft-parameters-2} give material-specific PAW datasets, plane-wave cutoffs and displacement counts. FC2/FC3 supercells are primitive-cell multiples. Phono3py 2.3.2/phonopy 2.15.1 transport uses Wigner/RTA and total population-plus-coherence trace averages~\cite{Phono3py2023,Simoncelli2022}, without NAC, four-phonon scattering or temperature-renormalized force constants. Harmonic-stability criteria are evaluated on the transport mesh, and small negative frequencies along denser paths are reported independently.
\begin{fixedtable}
\small
\caption{Shared settings for lattice-conductivity DFT validation.}
\label{tab:sm-common-dft}
\begin{tabular}{p{.29\textwidth}p{.65\textwidth}}\toprule
Item & Setting \\\midrule
Electronic structure & VASP 6.5.0; PAW--PBE; ISPIN=1; no SOC or added Hubbard correction \\
Cutoff & $1.3\max(\mathrm{ENMAX})$; 286.1456--525.1077 eV \\
Electronic convergence & EDIFF=$10^{-8}$ eV; NELM=200; NELMIN=5; ALGO=Normal \\
Accuracy/occupations & PREC=Accurate; LREAL=False; ADDGRID=True; LASPH=True; ISMEAR=0; SIGMA=0.01 eV \\
Optimization & IBRION=2; ISIF=3; PSTRESS=0; up to 10 ionic steps per segment and 200 ionic steps overall; POTIM=0.2; EDIFFG=$-10^{-4}$ eV/\AA \\
Final self-consistent static & ISTART=0; ICHARG=2; IBRION=$-1$; NSW=0; ISIF=2 \\
Final structural convergence criteria & Force $\leq10^{-3}$ eV/\AA; maximum absolute stress $\leq0.05$ GPa; energy increase $\leq0.01$ eV/atom \\
Symmetry & Optimization/static: ISYM=2, SYMPREC=$10^{-5}$ \AA; displaced forces: ISYM=0 \\
Electronic meshes & Gamma-centered $6^3$ for primitive optimization/static, $2^3$ for FC2, $3^3$ for FC3; zero shift \\
Force constants & Primitive-cell FC2 $4^3$/FC3 $2^3$; 0.03 \AA{} displacement; real-space symmetrization \\
Transport & 300 K; $19^3$ mesh; tetrahedron integration; Wigner/RTA; isotope scattering; 1 m boundary; no NAC \\
Scalar output & Total population-plus-coherence $(\kappa_{xx}+\kappa_{yy}+\kappa_{zz})/3$; no electronic conductivity \\
Computational configuration & VASP: 64 MPI ranks, one thread/rank, NCORE=8; transport calculations were configured to use 64 OpenMP threads \\\bottomrule
\end{tabular}
\end{fixedtable}
\begin{fixedtable}
\centering
\small
\setlength{\tabcolsep}{3pt}
\caption{Material-specific DFT parameters (part 1 of 2). $N$ is the primitive-cell atom count; $n_2/n_3$ are the actual independent FC2/FC3 force calculations. PAW datasets are listed in atomic-species order.}
\begin{tabular}{lrrrlrr}
\hline
Material & SG & $N$ & $E_{\rm cut}$ (eV) & PAW datasets & $n_2$ & $n_3$ \\
\hline
$\mathrm{CsKAu_{2}}$ & 225 & 4 & 337.0432 & Au, Cs\_sv, K\_sv & 3 & 231 \\
$\mathrm{SrLaPbAu}$ & 216 & 4 & 309.1855 & Au, La, Pb\_d, Sr\_sv & 4 & 452 \\
$\mathrm{CsBaHgSb}$ & 216 & 4 & 303.1652 & Ba\_sv, Cs\_sv, Hg, Sb & 4 & 452 \\
$\mathrm{CsKAuBr}$ & 216 & 4 & 337.0432 & Au, Br, Cs\_sv, K\_sv & 4 & 452 \\
$\mathrm{CsKRbSb}$ & 216 & 4 & 337.0432 & Cs\_sv, K\_sv, Rb\_sv, Sb & 4 & 452 \\
$\mathrm{CsRbAu_{2}}$ & 225 & 4 & 298.9259 & Au, Cs\_sv, Rb\_sv & 3 & 231 \\
$\mathrm{KNaHgTe}$ & 216 & 4 & 337.4293 & Hg, K\_sv, Na\_pv, Te & 4 & 452 \\
$\mathrm{CsKNaSb}$ & 216 & 4 & 337.4293 & Cs\_sv, K\_sv, Na\_pv, Sb & 4 & 452 \\
$\mathrm{Rb_{2}HgS}$ & 225 & 4 & 336.2957 & Hg, Rb\_sv, S & 3 & 231 \\
$\mathrm{NaTlHgTe}$ & 216 & 4 & 337.4293 & Hg, Na\_pv, Te, Tl\_d & 4 & 452 \\
$\mathrm{K_{2}HgSe}$ & 225 & 4 & 337.0432 & Hg, K\_sv, Se & 3 & 231 \\
$\mathrm{Cs_{2}AuI}$ & 225 & 4 & 298.9259 & Au, Cs\_sv, I & 3 & 231 \\
$\mathrm{CaLaPbAu}$ & 216 & 4 & 346.6086 & Au, Ca\_sv, La, Pb\_d & 4 & 452 \\
$\mathrm{SrLaSnAu}$ & 216 & 4 & 313.4079 & Au, La, Sn\_d, Sr\_sv & 4 & 452 \\
$\mathrm{CsRbAuCl}$ & 216 & 4 & 341.2136 & Au, Cl, Cs\_sv, Rb\_sv & 4 & 452 \\
$\mathrm{Cs_{2}AuCl}$ & 225 & 4 & 341.2136 & Au, Cl, Cs\_sv & 3 & 231 \\
$\mathrm{RbSrHgBi}$ & 216 & 4 & 315.6907 & Bi\_d, Hg, Rb\_sv, Sr\_sv & 4 & 452 \\
$\mathrm{RbBaHgSb}$ & 216 & 4 & 303.1652 & Ba\_sv, Hg, Rb\_sv, Sb & 4 & 452 \\
$\mathrm{KRbI_{2}}$ & 225 & 4 & 337.0432 & I, K\_sv, Rb\_sv & 3 & 231 \\
$\mathrm{RbSrHgSb}$ & 216 & 4 & 303.1652 & Hg, Rb\_sv, Sb, Sr\_sv & 4 & 452 \\
$\mathrm{Rb_{2}AuBr}$ & 225 & 4 & 298.9259 & Au, Br, Rb\_sv & 3 & 231 \\
$\mathrm{BaNaHgSb}$ & 216 & 4 & 337.4293 & Ba\_sv, Hg, Na\_pv, Sb & 4 & 452 \\
$\mathrm{NaHgAsPb}$ & 216 & 4 & 337.4293 & As, Hg, Na\_pv, Pb\_d & 4 & 452 \\
$\mathrm{Cs_{2}AuBr}$ & 225 & 4 & 298.9259 & Au, Br, Cs\_sv & 3 & 231 \\
$\mathrm{Cs_{2}AgBr}$ & 225 & 4 & 324.7972 & Ag, Br, Cs\_sv & 3 & 231 \\
$\mathrm{KNaTlBi}$ & 216 & 4 & 337.4293 & Bi\_d, K\_sv, Na\_pv, Tl\_d & 4 & 452 \\
$\mathrm{MgTlBi}$ & 216 & 3 & 315.6907 & Bi\_d, Mg, Tl\_d & 3 & 255 \\
$\mathrm{NaSnHgSb}$ & 216 & 4 & 337.4293 & Hg, Na\_pv, Sb, Sn\_d & 4 & 452 \\
$\mathrm{PbCl_{2}}$ & 225 & 3 & 341.2136 & Cl, Pb\_d & 2 & 128 \\
\hline
\end{tabular}
\label{tab:sm-dft-parameters-1}
\end{fixedtable}

\begin{fixedtable}
\centering
\small
\setlength{\tabcolsep}{3pt}
\caption{Material-specific DFT parameters (part 2 of 2). $N$ is the primitive-cell atom count; $n_2/n_3$ are the actual independent FC2/FC3 force calculations. PAW datasets are listed in atomic-species order.}
\begin{tabular}{lrrrlrr}
\hline
Material & SG & $N$ & $E_{\rm cut}$ (eV) & PAW datasets & $n_2$ & $n_3$ \\
\hline
$\mathrm{CaInGaHg}$ & 216 & 4 & 367.4983 & Ca\_sv, Ga\_d, Hg, In\_d & 4 & 452 \\
$\mathrm{KBaHgBi}$ & 216 & 4 & 337.0432 & Ba\_sv, Bi\_d, Hg, K\_sv & 4 & 452 \\
$\mathrm{RbAu}$ & 187 & 2 & 298.9259 & Au, Rb\_sv & 4 & 164 \\
$\mathrm{KBaCdBi}$ & 216 & 4 & 356.6368 & Ba\_sv, Bi\_d, Cd, K\_sv & 4 & 452 \\
$\mathrm{BaCaHgGe}$ & 216 & 4 & 403.3822 & Ba\_sv, Ca\_sv, Ge\_d, Hg & 4 & 452 \\
$\mathrm{CaTlCdIn}$ & 216 & 4 & 356.6368 & Ca\_sv, Cd, In\_d, Tl\_d & 4 & 452 \\
$\mathrm{CsRbCl_{2}}$ & 225 & 4 & 341.2136 & Cl, Cs\_sv, Rb\_sv & 3 & 231 \\
$\mathrm{LuMgSnAu}$ & 216 & 4 & 525.1077 & Au, Lu\_3, Mg\_pv, Sn\_d & 4 & 452 \\
$\mathrm{KCaHgSb}$ & 216 & 4 & 346.6086 & Ca\_sv, Hg, K\_sv, Sb & 4 & 452 \\
$\mathrm{KBaNaSn}$ & 216 & 4 & 337.4293 & Ba\_sv, K\_sv, Na\_pv, Sn\_d & 4 & 452 \\
$\mathrm{CsKS}$ & 216 & 3 & 337.0432 & Cs\_sv, K\_sv, S & 3 & 255 \\
$\mathrm{KNa_{2}As}$ & 225 & 4 & 337.4293 & As, K\_sv, Na\_pv & 3 & 231 \\
$\mathrm{NaLaBiAu}$ & 216 & 4 & 337.4293 & Au, Bi\_d, La, Na\_pv & 4 & 452 \\
$\mathrm{Rb_{2}Pd}$ & 225 & 3 & 326.2025 & Pd, Rb\_sv & 2 & 128 \\
$\mathrm{BaBrCl}$ & 216 & 3 & 341.2136 & Ba\_sv, Br, Cl & 3 & 255 \\
$\mathrm{RbTlS}$ & 156 & 3 & 336.2957 & Rb\_sv, S, Tl\_d & 9 & 657 \\
$\mathrm{BaPd}$ & 221 & 2 & 326.2025 & Ba\_sv, Pd & 2 & 42 \\
$\mathrm{TlBiPb}$ & 216 & 3 & 315.6907 & Bi\_d, Pb\_d, Tl\_d & 3 & 255 \\
$\mathrm{CsAg}$ & 225 & 2 & 324.7972 & Ag, Cs\_sv & 2 & 62 \\
$\mathrm{KNa_{2}Bi}$ & 225 & 4 & 337.4293 & Bi\_d, K\_sv, Na\_pv & 3 & 231 \\
$\mathrm{NaRbS}$ & 187 & 3 & 337.4293 & Na\_pv, Rb\_sv, S & 6 & 370 \\
$\mathrm{RbBaBi}$ & 216 & 3 & 315.6907 & Ba\_sv, Bi\_d, Rb\_sv & 3 & 255 \\
$\mathrm{CsAgS}$ & 187 & 3 & 336.2957 & Ag, Cs\_sv, S & 6 & 370 \\
$\mathrm{CsAgSe}$ & 187 & 3 & 324.7972 & Ag, Cs\_sv, Se & 6 & 370 \\
$\mathrm{RbAgS}$ & 187 & 3 & 336.2957 & Ag, Rb\_sv, S & 6 & 370 \\
$\mathrm{SrIBr}$ & 156 & 3 & 298.1589 & Br, I, Sr\_sv & 9 & 657 \\
$\mathrm{RbBaAs}$ & 216 & 3 & 286.1456 & As, Ba\_sv, Rb\_sv & 3 & 255 \\
$\mathrm{Sr_{2}ZnSi}$ & 225 & 4 & 359.7399 & Si, Sr\_sv, Zn & 3 & 231 \\
$\mathrm{BaTlP}$ & 156 & 3 & 331.5520 & Ba\_sv, P, Tl\_d & 9 & 657 \\
\hline
\end{tabular}
\label{tab:sm-dft-parameters-2}
\end{fixedtable}

The total wall-clock computation time averages 24.530 h per structure (0.831--90.150 h), excluding queue waiting. It includes optimization, force evaluations, transport calculations, and unsuccessful or repeated calculations. The VASP time in Table~\ref{tab:sm-dft-times} is a subset of this total. Later supplementary temperature postprocessing is excluded. Table~\ref{tab:sm-dft-times} reports the total and VASP times as two separate quantities.

\begin{fixedtable}
\small\setlength{\tabcolsep}{3pt}
\caption{Total and successful VASP times in hours. Total time is cumulative wall-clock computation time and includes unsuccessful or repeated calculations; VASP time is a subset.}
\label{tab:sm-dft-times}
\begin{tabular*}{\textwidth}{@{\extracolsep{\fill}}lrr@{\hspace{15pt}}lrr@{}}
\toprule Material & Total time (h) & VASP & Material & Total time (h) & VASP \\\midrule
$\mathrm{CsKAu_{2}}$ & 21.163 & 21.129 & $\mathrm{CaInGaHg}$ & 31.714 & 31.629 \\
$\mathrm{SrLaPbAu}$ & 61.891 & 55.798 & $\mathrm{KBaHgBi}$ & 52.369 & 52.291 \\
$\mathrm{CsBaHgSb}$ & 29.451 & 29.383 & $\mathrm{RbAu}$ & 2.221 & 2.206 \\
$\mathrm{CsKAuBr}$ & 27.942 & 27.858 & $\mathrm{KBaCdBi}$ & 63.288 & 57.202 \\
$\mathrm{CsKRbSb}$ & 49.786 & 49.691 & $\mathrm{BaCaHgGe}$ & 74.271 & 68.194 \\
$\mathrm{CsRbAu_{2}}$ & 16.381 & 16.353 & $\mathrm{CaTlCdIn}$ & 42.312 & 36.238 \\
$\mathrm{KNaHgTe}$ & 19.687 & 19.613 & $\mathrm{CsRbCl_{2}}$ & 10.027 & 9.996 \\
$\mathrm{CsKNaSb}$ & 20.873 & 20.788 & $\mathrm{LuMgSnAu}$ & 90.150 & 84.072 \\
$\mathrm{Rb_{2}HgS}$ & 16.562 & 16.511 & $\mathrm{KCaHgSb}$ & 25.191 & 25.122 \\
$\mathrm{NaTlHgTe}$ & 23.152 & 23.087 & $\mathrm{KBaNaSn}$ & 33.297 & 33.231 \\
$\mathrm{K_{2}HgSe}$ & 14.547 & 14.518 & $\mathrm{CsKS}$ & 7.068 & 7.046 \\
$\mathrm{Cs_{2}AuI}$ & 20.872 & 20.833 & $\mathrm{KNa_{2}As}$ & 6.411 & 6.383 \\
$\mathrm{CaLaPbAu}$ & 77.463 & 77.392 & $\mathrm{NaLaBiAu}$ & 53.251 & 53.184 \\
$\mathrm{SrLaSnAu}$ & 54.640 & 54.557 & $\mathrm{Rb_{2}Pd}$ & 9.140 & 9.115 \\
$\mathrm{CsRbAuCl}$ & 25.653 & 25.568 & $\mathrm{BaBrCl}$ & 8.704 & 8.681 \\
$\mathrm{Cs_{2}AuCl}$ & 19.809 & 19.777 & $\mathrm{RbTlS}$ & 21.065 & 21.005 \\
$\mathrm{RbSrHgBi}$ & 46.141 & 40.070 & $\mathrm{BaPd}$ & 0.831 & 0.802 \\
$\mathrm{RbBaHgSb}$ & 24.456 & 24.381 & $\mathrm{TlBiPb}$ & 9.815 & 9.794 \\
$\mathrm{KRbI_{2}}$ & 21.303 & 21.274 & $\mathrm{CsAg}$ & 1.684 & 1.667 \\
$\mathrm{RbSrHgSb}$ & 22.889 & 22.823 & $\mathrm{KNa_{2}Bi}$ & 11.906 & 11.880 \\
$\mathrm{Rb_{2}AuBr}$ & 11.751 & 11.723 & $\mathrm{NaRbS}$ & 6.716 & 6.690 \\
$\mathrm{BaNaHgSb}$ & 22.741 & 22.659 & $\mathrm{RbBaBi}$ & 20.204 & 20.180 \\
$\mathrm{NaHgAsPb}$ & 14.947 & 14.903 & $\mathrm{CsAgS}$ & 7.276 & 7.247 \\
$\mathrm{Cs_{2}AuBr}$ & 14.206 & 14.173 & $\mathrm{CsAgSe}$ & 7.419 & 7.391 \\
$\mathrm{Cs_{2}AgBr}$ & 23.304 & 23.264 & $\mathrm{RbAgS}$ & 7.143 & 7.112 \\
$\mathrm{KNaTlBi}$ & 30.862 & 30.781 & $\mathrm{SrIBr}$ & 20.312 & 20.238 \\
$\mathrm{MgTlBi}$ & 4.859 & 4.841 & $\mathrm{RbBaAs}$ & 5.407 & 5.381 \\
$\mathrm{NaSnHgSb}$ & 17.357 & 17.319 & $\mathrm{Sr_{2}ZnSi}$ & 19.846 & 19.813 \\
$\mathrm{PbCl_{2}}$ & 2.891 & 2.867 & $\mathrm{BaTlP}$ & 16.115 & 16.056 \\
\bottomrule\end{tabular*}
\end{fixedtable}

\section{Empirical computational-budget filter}\label{app:computational-budget}

Figure~\ref{fig:Fig-screening-two_thresholds-2-and-8} presents the full-database prescreening, which uses 1,953,259 initial database structures as input. The DANA-$\kappa$ predictions on these initial structures were partitioned into two non-overlapping branches: a priority branch ($\hat{\kappa}_L<2~\mathrm{W\,m^{-1}\,K^{-1}}$) and an extended branch ($2\le\hat{\kappa}_L<8~\mathrm{W\,m^{-1}\,K^{-1}}$), yielding 656,693 and 970,442 records, respectively; records with $\hat{\kappa}_L\ge8~\mathrm{W\,m^{-1}\,K^{-1}}$ were not assigned to either branch. After structural, electronic, and thermodynamic filters---a primitive-cell size of no more than 20 atoms, a band gap greater than 0.1~eV, nonmagnetic ordering, and an energy above the convex hull of at most 100~meV/atom---29,177 and 18,411 records were retained in the two branches. Intersecting these with the original cost-selected candidate pool yielded 1,735 and 2,785 records, of which 1,324 and 2,086 had available GRACE thermal conductivities; respectively, 1,167 and 526 records satisfied $0<\kappa_{\mathrm{GRACE}}<2~\mathrm{W\,m^{-1}\,K^{-1}}$ and were combined into a GRACE low-thermal-conductivity pool of 1,693 records. A uniform DFT protocol was then completed for 59 candidates, of which 58 satisfy $\kappa_{\mathrm{DFT}}<2~\mathrm{W\,m^{-1}\,K^{-1}}$, including 50 from the priority branch and 8 from the extended branch; 29 of these are below $0.5~\mathrm{W\,m^{-1}\,K^{-1}}$. The extended branch primarily serves to recover low-thermal-conductivity candidates missed by the strict threshold because of DANA-$\kappa$ overestimation. Its GRACE low-thermal-conductivity hit rate is markedly lower than that of the priority branch, making it suitable as a recall-recovery step rather than an independent, efficient priority-screening strategy.

The two-threshold branches contain 1,693 GRACE low-thermal-conductivity records, compared with 1,712 in the full GRACE low-thermal-conductivity dataset. The difference of 19 records corresponds to structures with DANA-$\kappa$ predictions $\ge8~\mathrm{W\,m^{-1}\,K^{-1}}$ on the initial structures, which were not assigned to either branch in this figure. In addition, the DANA predictions in Table~\ref{tab:dft-low-kappa} were evaluated a posteriori on DFT-relaxed structures, whereas the branch assignment in this figure is based on the initial database structures; the two therefore involve different structural configurations. Consequently, the nine materials with DANA predictions above 2 in Table~\ref{tab:dft-low-kappa} do not necessarily fall outside the initial screening: seven belong to the extended branch; BaBrCl and RbBaAs were below 2 at the initial screening stage but increased after relaxation; and RbBaBi, another material from the extended branch, decreased to 1.9531 after relaxation. Overall, DANA-$\kappa$ is best positioned as a tool for candidate prioritization and rapid prescreening rather than as a final criterion for low thermal conductivity, and final conclusions should still be drawn from GRACE transport calculations and DFT validation.

\begin{fixedfigure}
\includegraphics[width=\textwidth,height=.85\textheight,keepaspectratio]{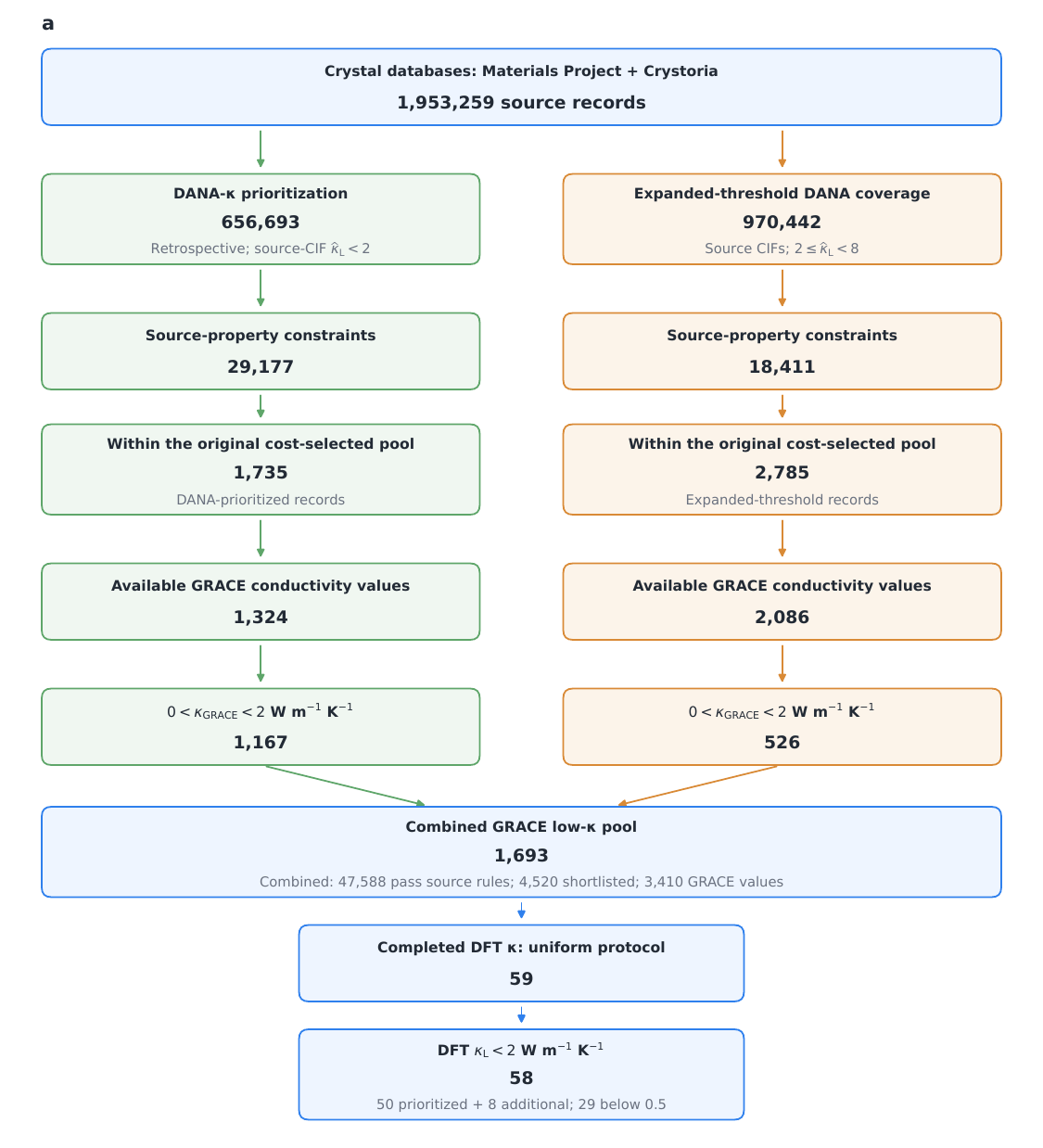}
\caption{Screening workflow based on DANA-$\kappa$ prediction thresholds. The 1,953,259 initial database structures were divided according to their DANA-$\kappa$ predictions into a priority branch ($\hat{\kappa}_L<2~\mathrm{W\,m^{-1}\,K^{-1}}$) and an extended branch ($2\le\hat{\kappa}_L<8~\mathrm{W\,m^{-1}\,K^{-1}}$). After structural, electronic, and thermodynamic constraints, intersection with the original cost-selected candidate pool, and GRACE transport evaluation, the two branches were combined into 1,693 GRACE low-thermal-conductivity records. A uniform DFT protocol was then completed for 59 candidates, of which 58 satisfy $\kappa_{\mathrm{DFT}}<2~\mathrm{W\,m^{-1}\,K^{-1}}$, including 50 from the priority branch and 8 from the extended branch.}
\label{fig:Fig-screening-two_thresholds-2-and-8}
\end{fixedfigure}

Before force evaluations, an empirical estimate of computational cost used primitive-cell size and symmetry rather than explicitly generating every displacement dataset. Let $N_{\rm p}$ be the primitive atom count, $C$ the conventional-cell centring multiplicity, and $G$ the crystallographic point-group order. We used $C=1$ for P, 2 for A/B/C/I, 3 for R, and 4 for F cells. The screening proxy follows the conventional-cell GRACE workload,
\begin{equation}
 N_{\rm conv}=C N_{\rm p},\qquad
 n_2=64N_{\rm conv},\qquad n_3=8N_{\rm conv}.
 \label{eq:budget-supercells}
\end{equation}
The central estimates for independent force configurations are
\begin{equation}
 \widehat d_2=\left\lceil\frac{6N_{\rm p}}{G^{0.6}}\right\rceil,\qquad
 \widehat d_3=\left\lceil\frac{6N_{\rm p}(1+6n_3)}{G^{0.9}}\right\rceil.
 \label{eq:budget-displacements}
\end{equation}
The relative force-evaluation cost, memory required for a full third-order force-constant array in double precision, and relative transport cost are
\begin{align}
 W_{\rm rel}&=\frac{\widehat d_3n_3+\widehat d_2n_2}{146\cdot64+2\cdot512},\\
 M_3&=216n_3^3\ {\rm bytes},\\
 K_{\rm rel}&=\left(\frac{N_{\rm p}}{2}\right)^3\frac{48}{G_{\rm r}},
 \label{eq:budget-proxies}
\end{align}
where $G_{\rm r}$ includes nonmagnetic time-reversal reduction: $G_{\rm r}=G$ for centrosymmetric groups and $2G$ otherwise. The denominator in $W_{\rm rel}$ is a fixed TlBr reference computation used only for normalization, and the transport normalization is unity for the two-atom BN reference. These normalizations provide relative cost proxies rather than material-specific wall-time calibrations.

\begin{fixedtable}
\small
\caption{Simultaneous conditions defining the computational-cost threshold for candidate screening.}
\label{tab:sm-cost-budget}
\begin{tabular}{ll}\toprule
Proxy & Requirement \\\midrule
Estimated FC3 configurations & $\widehat d_3\leq2000$ \\
Full FC3 array & $M_3/2^{30}\leq4$ GiB \\
Relative force work & $W_{\rm rel}\leq100$ \\
Relative transport work & $K_{\rm rel}\leq1000$ \\\bottomrule
\end{tabular}
\end{fixedtable}

\section{Ablation}\label{app:Ablation}

Table~\ref{tab:ablation-summary} presents the ablation results for DANA-$\kappa$. Using the full model's equal-weight average MAE across the Random, Space group, and OOD test sets (0.677) as the baseline, removing the robust loss yields the largest increase in average error ($\Delta\mathrm{Avg}=+0.0827$) and degrades performance on all three test sets. Removing multi-stat or slot attention increases the average error by $+0.0261$ and $+0.0210$, respectively, with both degradations mainly occurring on the OOD set (increasing from 1.182 to 1.257 and 1.250, respectively).

\begin{fixedtable}
\small
\caption{
Ablation summary. The full DANA model is reported as the frozen DANA-$\kappa$ benchmark entry; the ablated variants use random-seed training.
$\Delta\mathrm{Avg}$ is computed as $\mathrm{Avg}_{\mathrm{ablated}}-\mathrm{Avg}_{\mathrm{full}}$; positive values indicate a worse average MAE after removing the design.
}
\label{tab:ablation-summary}
\begin{tabular}{@{}llrrrrr@{}}
\toprule
Design & Scheme & Random & Space group & OOD & Avg & $\Delta\mathrm{Avg}$ \\
\midrule
Full DANA & DANA-$\kappa$ (benchmark) & 0.370 & 0.480 & 1.182 & \textbf{0.677} & --- \\
w/o robust loss & A: MSE + sum + BN & 0.393805 & 0.499147 & 1.386004 & 0.759652 & $+0.0827$ \\
w/o multi-stat & D: Pseudo-Huber + sum + LN & 0.365840 & 0.486350 & 1.257063 & 0.703084 & $+0.0261$ \\
w/o slot attention & E: Pseudo-Huber + multi-stat + LN & 0.365502 & 0.478933 & 1.249669 & 0.698035 & $+0.0210$ \\
\bottomrule
\end{tabular}
\vspace{2pt}
\end{fixedtable}

\section{Phonon transport of the 58 low-conductivity phases}\label{app:all-transport}
Figures~\ref{fig:sm-transport-01}--\ref{fig:sm-transport-15} follow the material order of Table~\ref{tab:dft-low-kappa}. Each row shows DFT dispersions (blue lines), GRACE dispersions (orange open circles), fixed-force-constant temperature-dependent conductivities, root-mean-square velocities $|\mathbf v|/\sqrt3$, and three-phonon lifetimes $1/(4\pi\gamma_3)$ at 300 K. DFT and GRACE use their respective relaxed geometries and actual supercells. Scatter plots omit frequencies $\leq0.01$ THz and invalid values; their density is not a DOS. Medians are weighted by population-conductivity contributions. Asterisks retain RbAu's conditional acoustic acceptance; daggers mark the dense-path negative frequencies of RbTlS/BaTlP. Neither flag proves complete dynamical stability.

\begin{fixedfigure}
\includegraphics[width=\textwidth,height=.88\textheight,keepaspectratio,trim=0 4pt 0 20pt,clip]{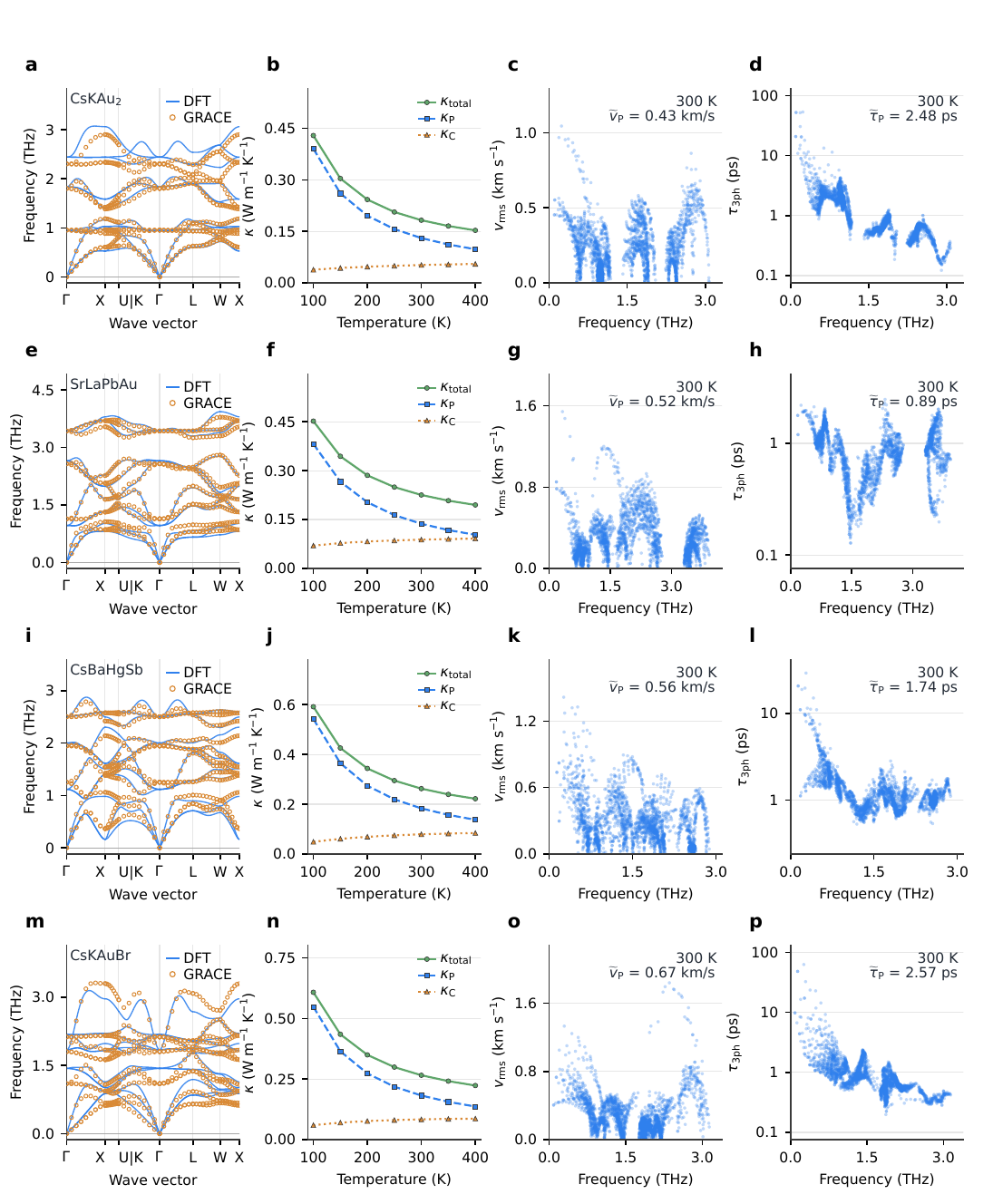}
\caption{Phonon and transport properties of $\mathrm{CsKAu_{2}}$, $\mathrm{SrLaPbAu}$, $\mathrm{CsBaHgSb}$, $\mathrm{CsKAuBr}$ (top to bottom). Columns show dispersions, total/population/coherence conductivity from 100--400 K, group velocity and three-phonon lifetime at 300 K. Definitions and quality markers are given above.}
\label{fig:sm-transport-01}
\end{fixedfigure}

\begin{fixedfigure}
\includegraphics[width=\textwidth,height=.88\textheight,keepaspectratio,trim=0 4pt 0 20pt,clip]{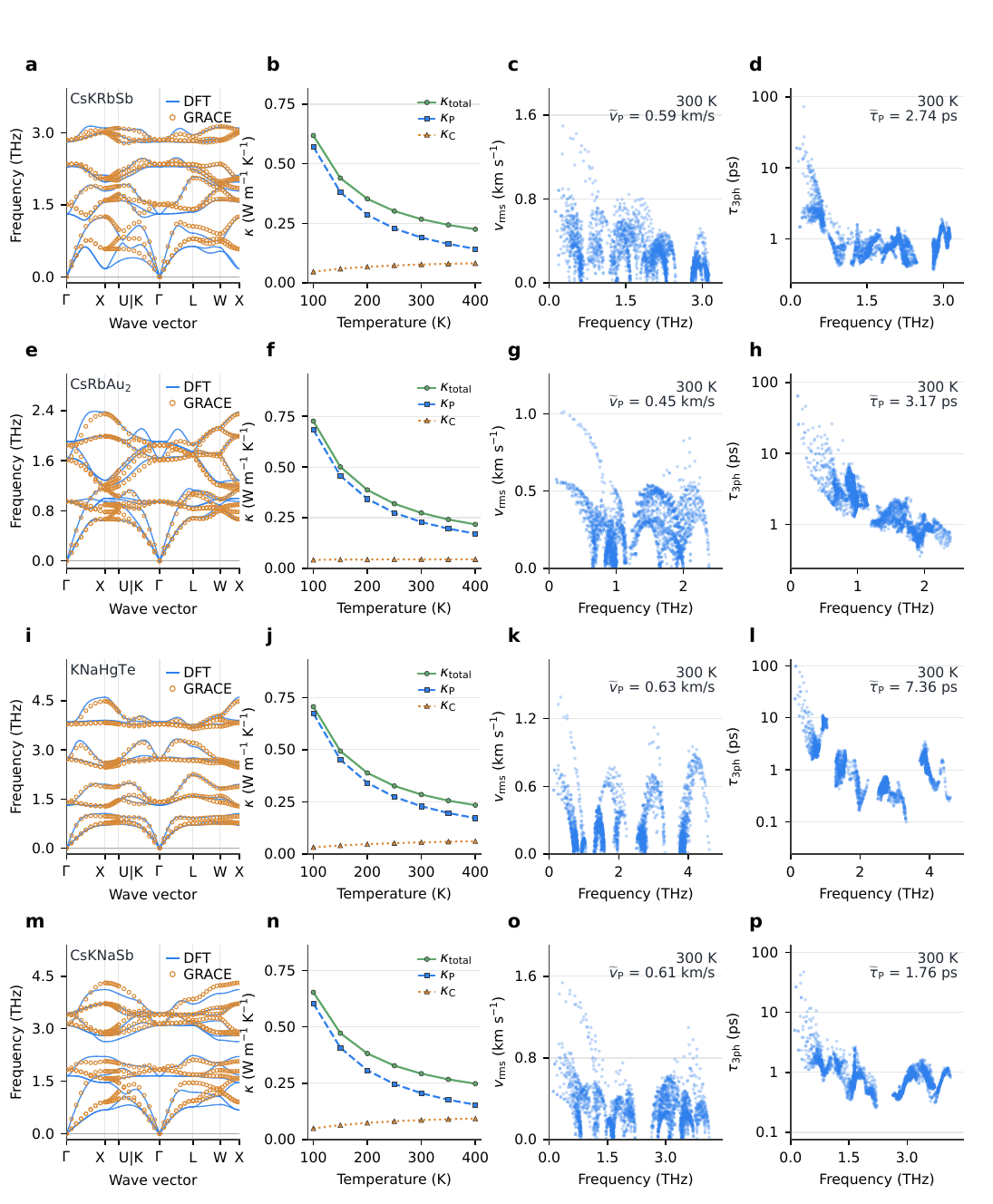}
\caption{Phonon and transport properties of $\mathrm{CsKRbSb}$, $\mathrm{CsRbAu_{2}}$, $\mathrm{KNaHgTe}$, $\mathrm{CsKNaSb}$ (top to bottom). Columns show dispersions, total/population/coherence conductivity from 100--400 K, group velocity and three-phonon lifetime at 300 K. Definitions and quality markers are given above.}
\label{fig:sm-transport-02}
\end{fixedfigure}

\begin{fixedfigure}
\includegraphics[width=\textwidth,height=.88\textheight,keepaspectratio,trim=0 4pt 0 20pt,clip]{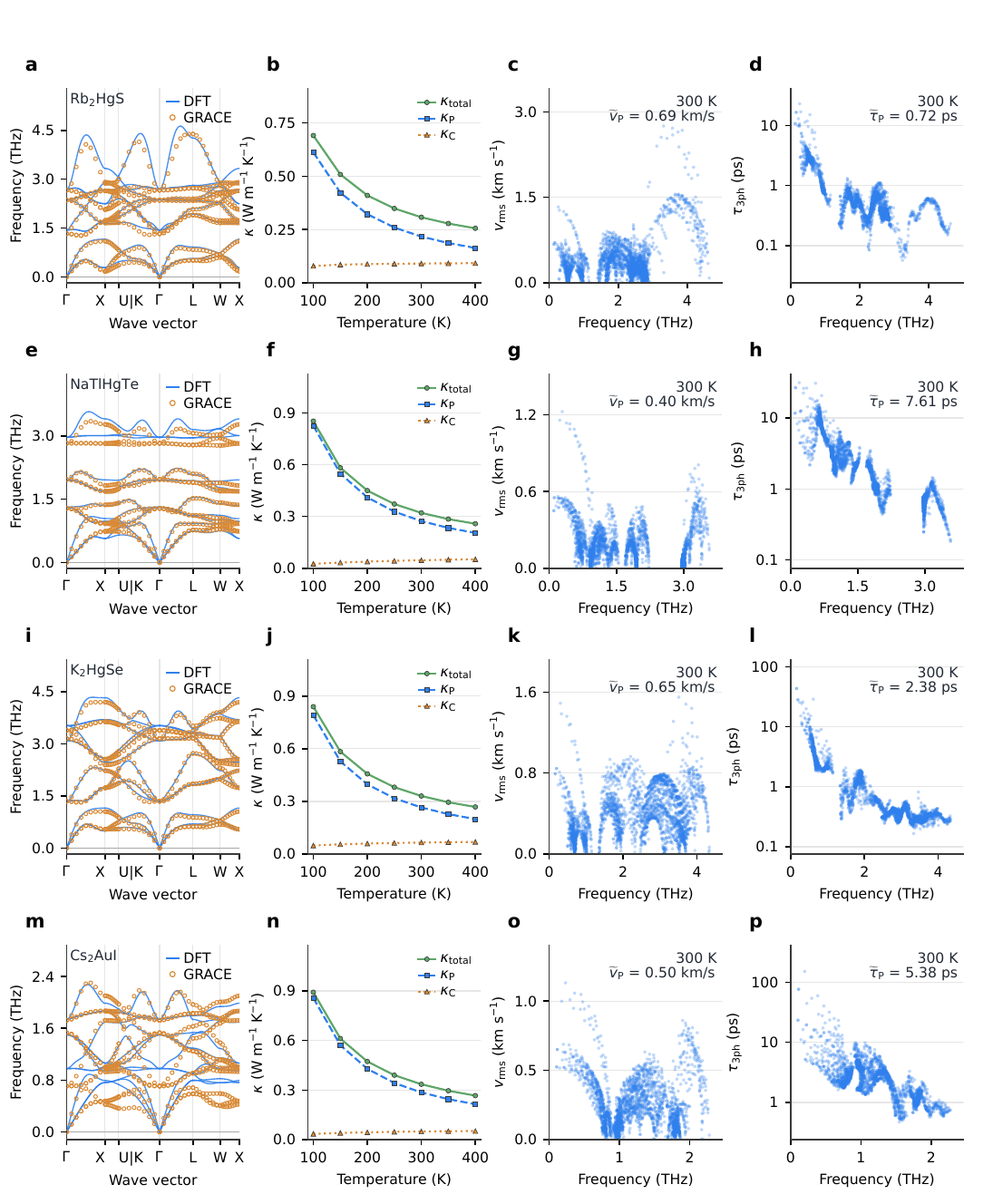}
\caption{Phonon and transport properties of $\mathrm{Rb_{2}HgS}$, $\mathrm{NaTlHgTe}$, $\mathrm{K_{2}HgSe}$, $\mathrm{Cs_{2}AuI}$ (top to bottom). Columns show dispersions, total/population/coherence conductivity from 100--400 K, group velocity and three-phonon lifetime at 300 K. Definitions and quality markers are given above.}
\label{fig:sm-transport-03}
\end{fixedfigure}

\begin{fixedfigure}
\includegraphics[width=\textwidth,height=.88\textheight,keepaspectratio,trim=0 4pt 0 20pt,clip]{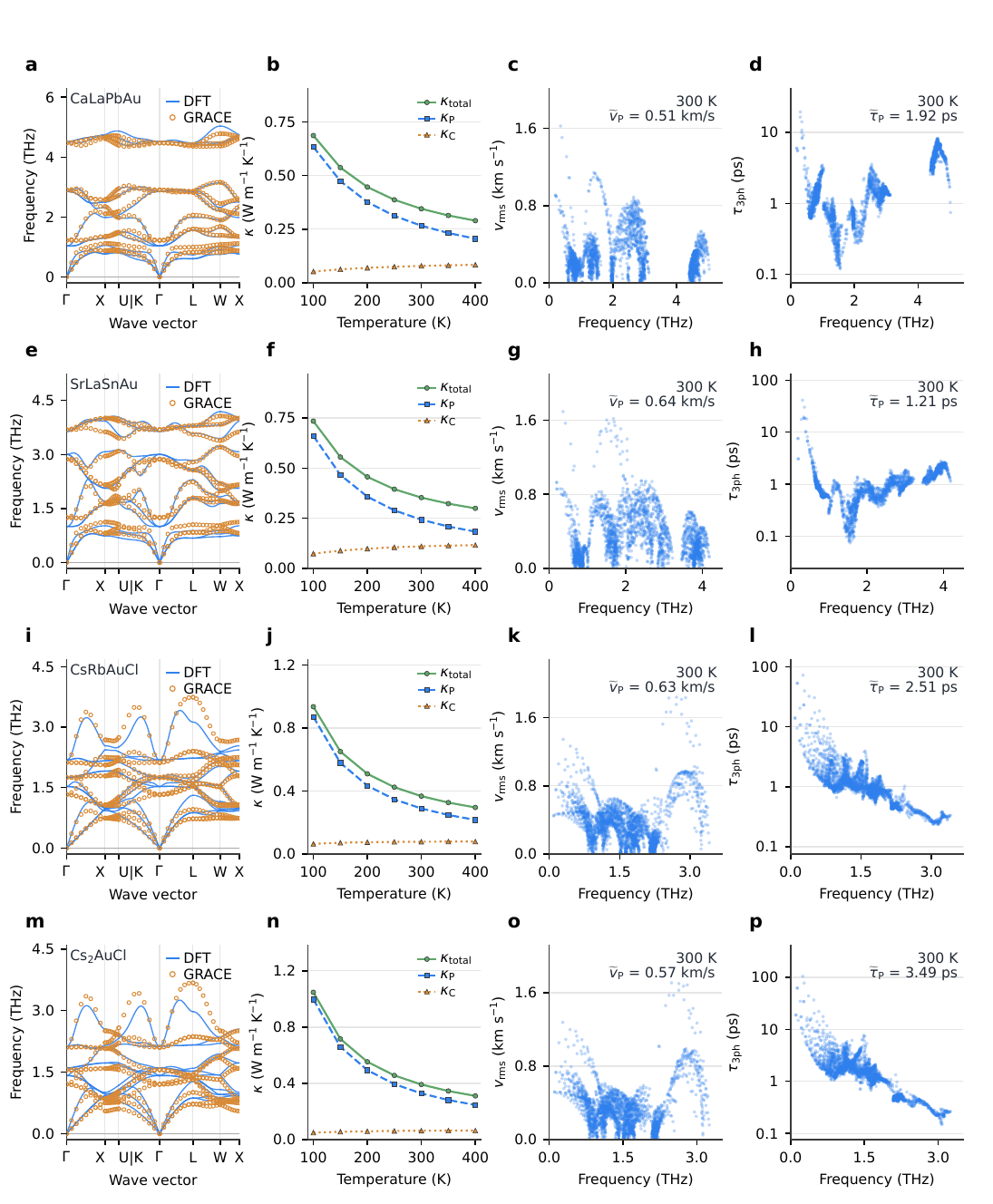}
\caption{Phonon and transport properties of $\mathrm{CaLaPbAu}$, $\mathrm{SrLaSnAu}$, $\mathrm{CsRbAuCl}$, $\mathrm{Cs_{2}AuCl}$ (top to bottom). Columns show dispersions, total/population/coherence conductivity from 100--400 K, group velocity and three-phonon lifetime at 300 K. Definitions and quality markers are given above.}
\label{fig:sm-transport-04}
\end{fixedfigure}

\begin{fixedfigure}
\includegraphics[width=\textwidth,height=.88\textheight,keepaspectratio,trim=0 4pt 0 20pt,clip]{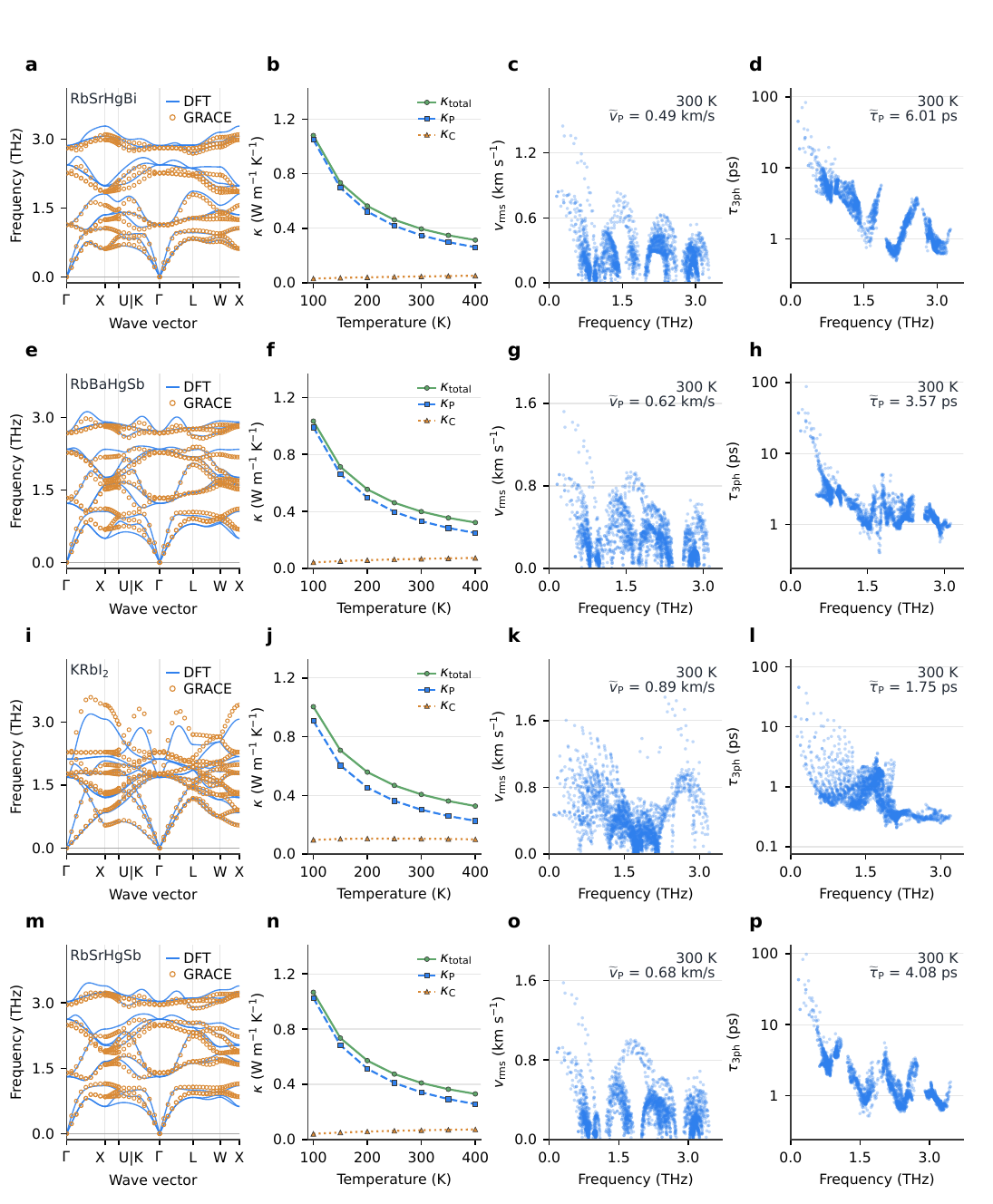}
\caption{Phonon and transport properties of $\mathrm{RbSrHgBi}$, $\mathrm{RbBaHgSb}$, $\mathrm{KRbI_{2}}$, $\mathrm{RbSrHgSb}$ (top to bottom). Columns show dispersions, total/population/coherence conductivity from 100--400 K, group velocity and three-phonon lifetime at 300 K. Definitions and quality markers are given above.}
\label{fig:sm-transport-05}
\end{fixedfigure}

\begin{fixedfigure}
\includegraphics[width=\textwidth,height=.88\textheight,keepaspectratio,trim=0 4pt 0 20pt,clip]{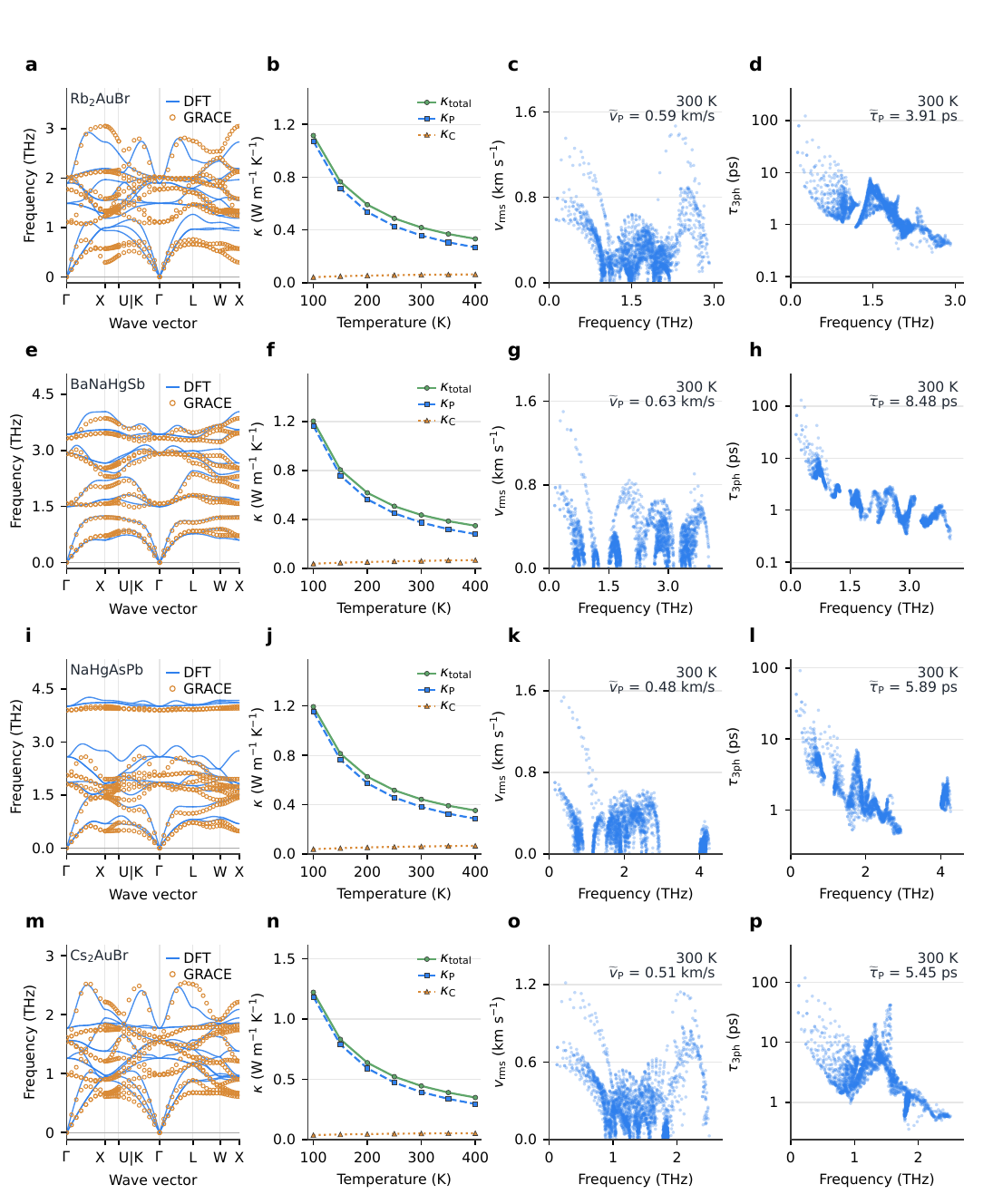}
\caption{Phonon and transport properties of $\mathrm{Rb_{2}AuBr}$, $\mathrm{BaNaHgSb}$, $\mathrm{NaHgAsPb}$, $\mathrm{Cs_{2}AuBr}$ (top to bottom). Columns show dispersions, total/population/coherence conductivity from 100--400 K, group velocity and three-phonon lifetime at 300 K. Definitions and quality markers are given above.}
\label{fig:sm-transport-06}
\end{fixedfigure}

\begin{fixedfigure}
\includegraphics[width=\textwidth,height=.88\textheight,keepaspectratio,trim=0 4pt 0 20pt,clip]{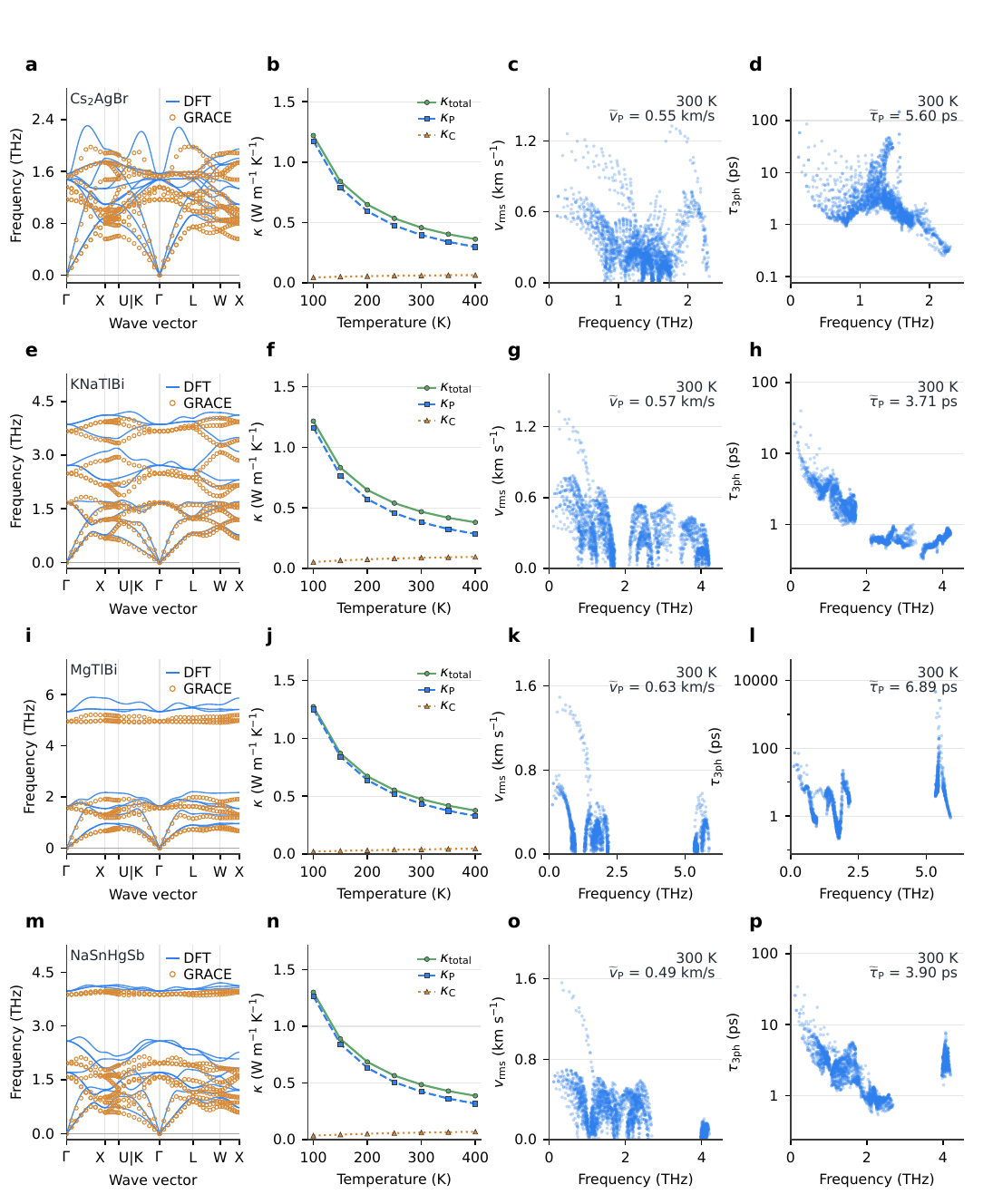}
\caption{Phonon and transport properties of $\mathrm{Cs_{2}AgBr}$, $\mathrm{KNaTlBi}$, $\mathrm{MgTlBi}$, $\mathrm{NaSnHgSb}$ (top to bottom). Columns show dispersions, total/population/coherence conductivity from 100--400 K, group velocity and three-phonon lifetime at 300 K. Definitions and quality markers are given above.}
\label{fig:sm-transport-07}
\end{fixedfigure}

\begin{fixedfigure}
\includegraphics[width=\textwidth,height=.88\textheight,keepaspectratio,trim=0 4pt 0 20pt,clip]{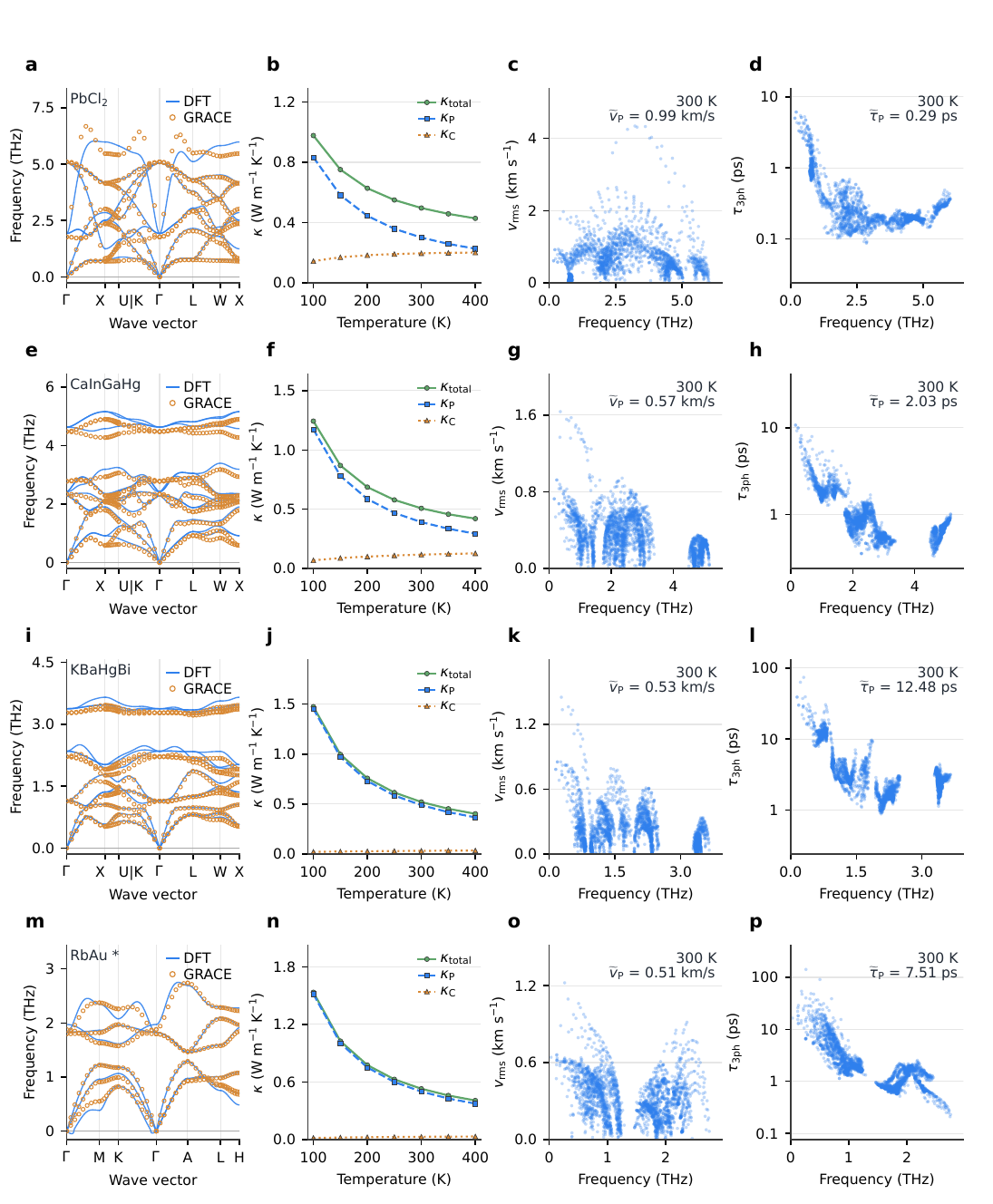}
\caption{Phonon and transport properties of $\mathrm{PbCl_{2}}$, $\mathrm{CaInGaHg}$, $\mathrm{KBaHgBi}$, $\mathrm{RbAu}$ (top to bottom). Columns show dispersions, total/population/coherence conductivity from 100--400 K, group velocity and three-phonon lifetime at 300 K. Definitions and quality markers are given above.}
\label{fig:sm-transport-08}
\end{fixedfigure}

\begin{fixedfigure}
\includegraphics[width=\textwidth,height=.88\textheight,keepaspectratio,trim=0 4pt 0 20pt,clip]{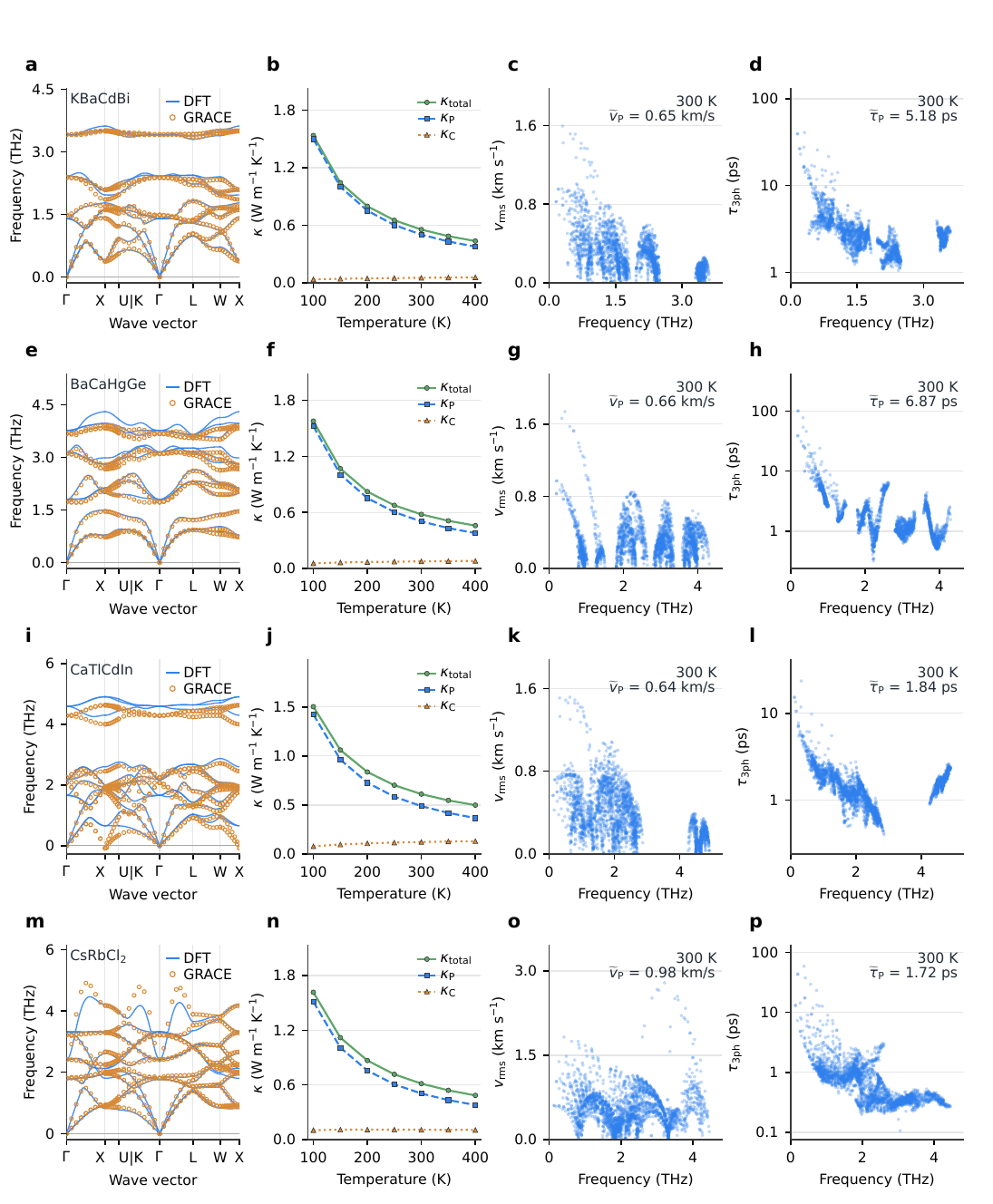}
\caption{Phonon and transport properties of $\mathrm{KBaCdBi}$, $\mathrm{BaCaHgGe}$, $\mathrm{CaTlCdIn}$, $\mathrm{CsRbCl_{2}}$ (top to bottom). Columns show dispersions, total/population/coherence conductivity from 100--400 K, group velocity and three-phonon lifetime at 300 K. Definitions and quality markers are given above.}
\label{fig:sm-transport-09}
\end{fixedfigure}

\begin{fixedfigure}
\includegraphics[width=\textwidth,height=.88\textheight,keepaspectratio,trim=0 4pt 0 20pt,clip]{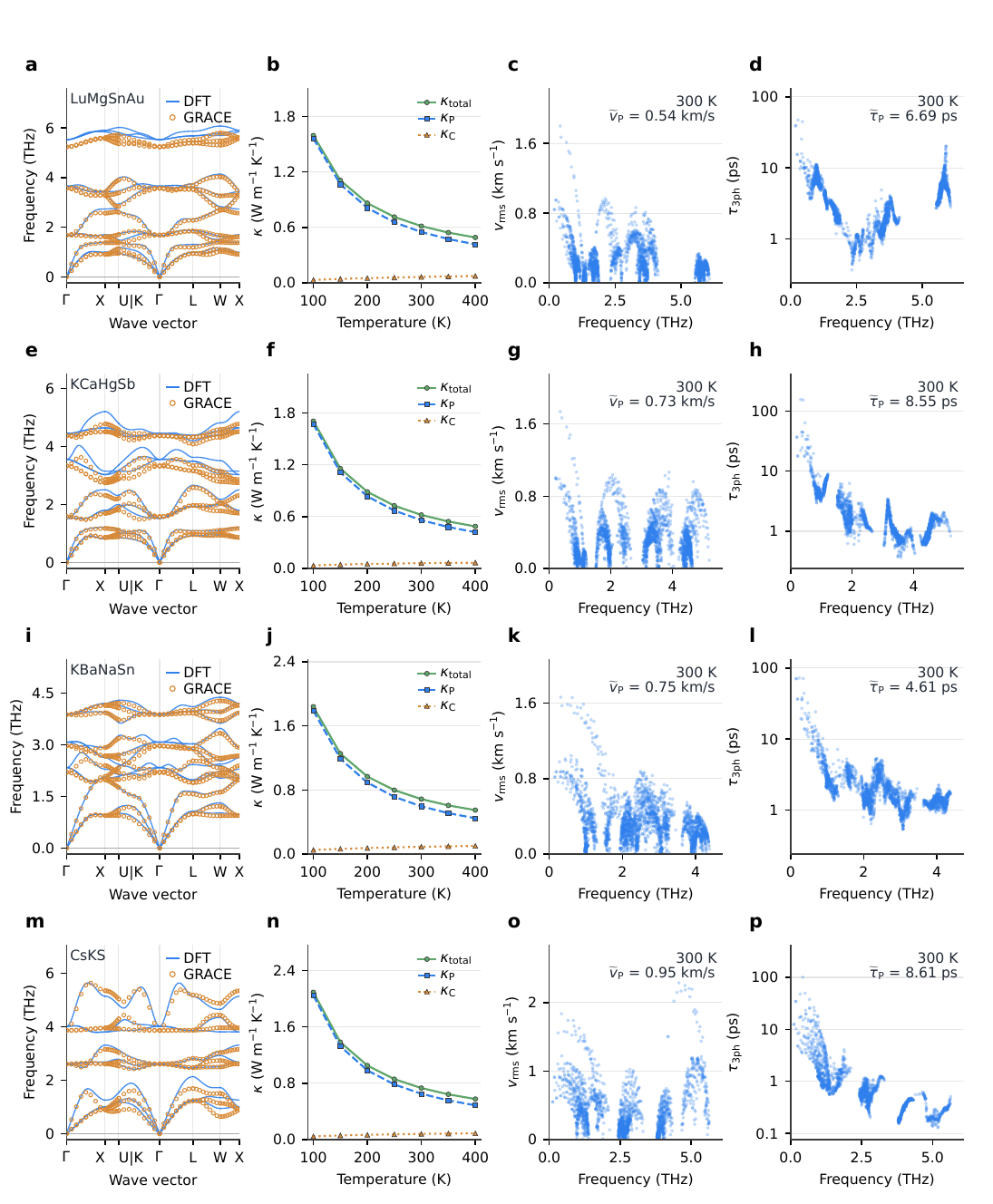}
\caption{Phonon and transport properties of $\mathrm{LuMgSnAu}$, $\mathrm{KCaHgSb}$, $\mathrm{KBaNaSn}$, $\mathrm{CsKS}$ (top to bottom). Columns show dispersions, total/population/coherence conductivity from 100--400 K, group velocity and three-phonon lifetime at 300 K. Definitions and quality markers are given above.}
\label{fig:sm-transport-10}
\end{fixedfigure}

\begin{fixedfigure}
\includegraphics[width=\textwidth,height=.88\textheight,keepaspectratio,trim=0 4pt 0 20pt,clip]{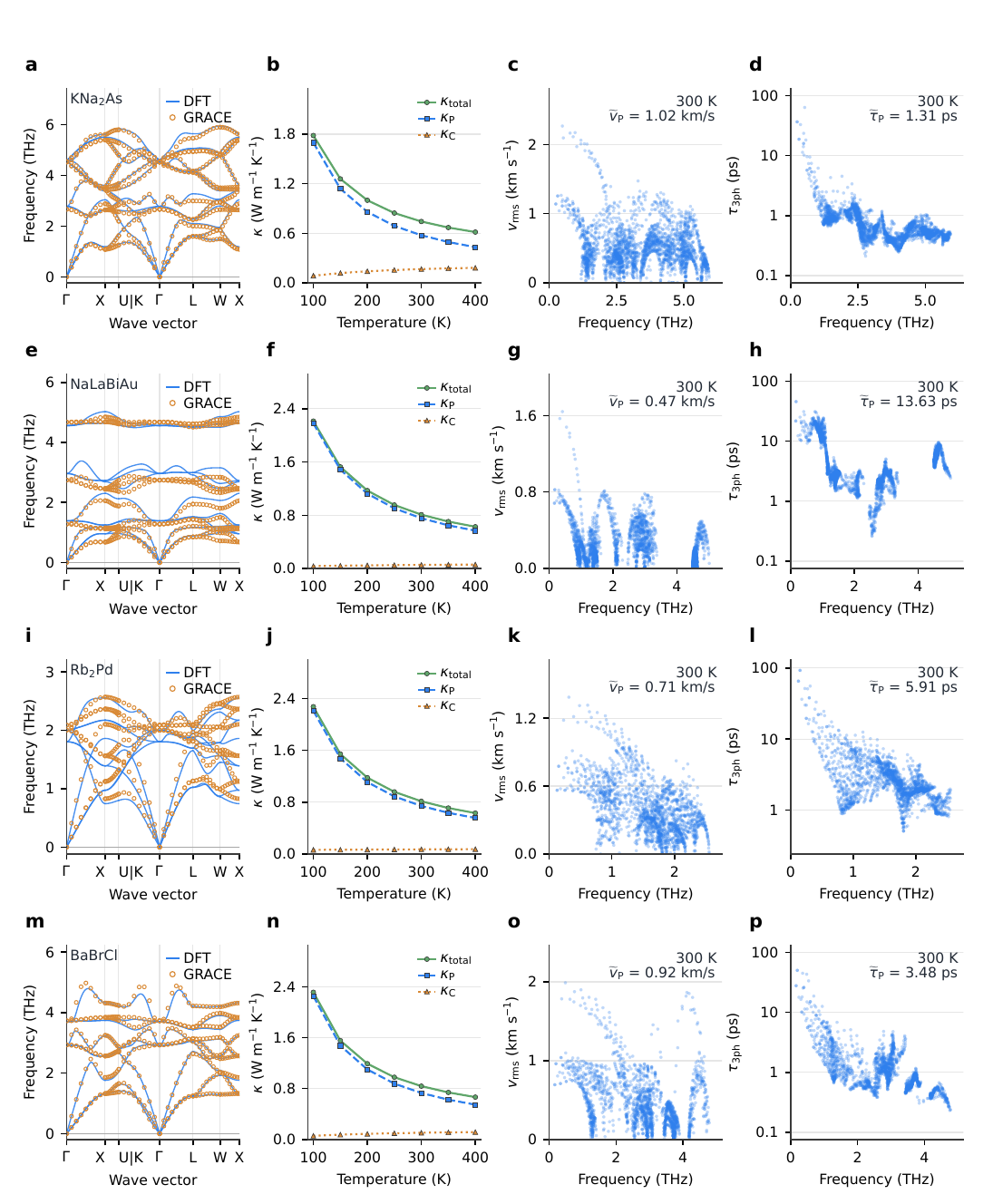}
\caption{Phonon and transport properties of $\mathrm{KNa_{2}As}$, $\mathrm{NaLaBiAu}$, $\mathrm{Rb_{2}Pd}$, $\mathrm{BaBrCl}$ (top to bottom). Columns show dispersions, total/population/coherence conductivity from 100--400 K, group velocity and three-phonon lifetime at 300 K. Definitions and quality markers are given above.}
\label{fig:sm-transport-11}
\end{fixedfigure}

\begin{fixedfigure}
\includegraphics[width=\textwidth,height=.88\textheight,keepaspectratio,trim=0 4pt 0 20pt,clip]{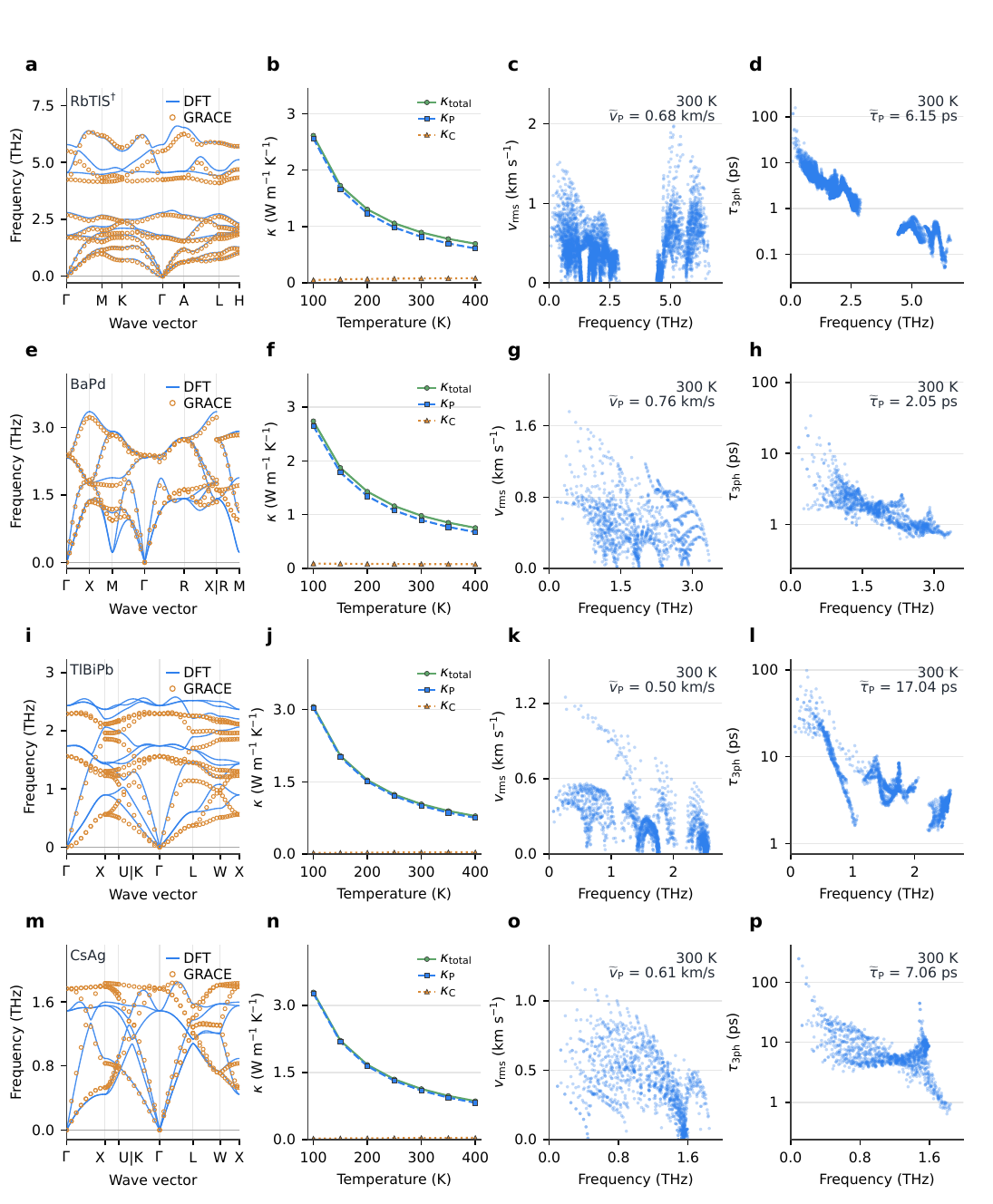}
\caption{Phonon and transport properties of $\mathrm{RbTlS}$, $\mathrm{BaPd}$, $\mathrm{TlBiPb}$, $\mathrm{CsAg}$ (top to bottom). Columns show dispersions, total/population/coherence conductivity from 100--400 K, group velocity and three-phonon lifetime at 300 K. Definitions and quality markers are given above.}
\label{fig:sm-transport-12}
\end{fixedfigure}

\begin{fixedfigure}
\includegraphics[width=\textwidth,height=.88\textheight,keepaspectratio,trim=0 4pt 0 20pt,clip]{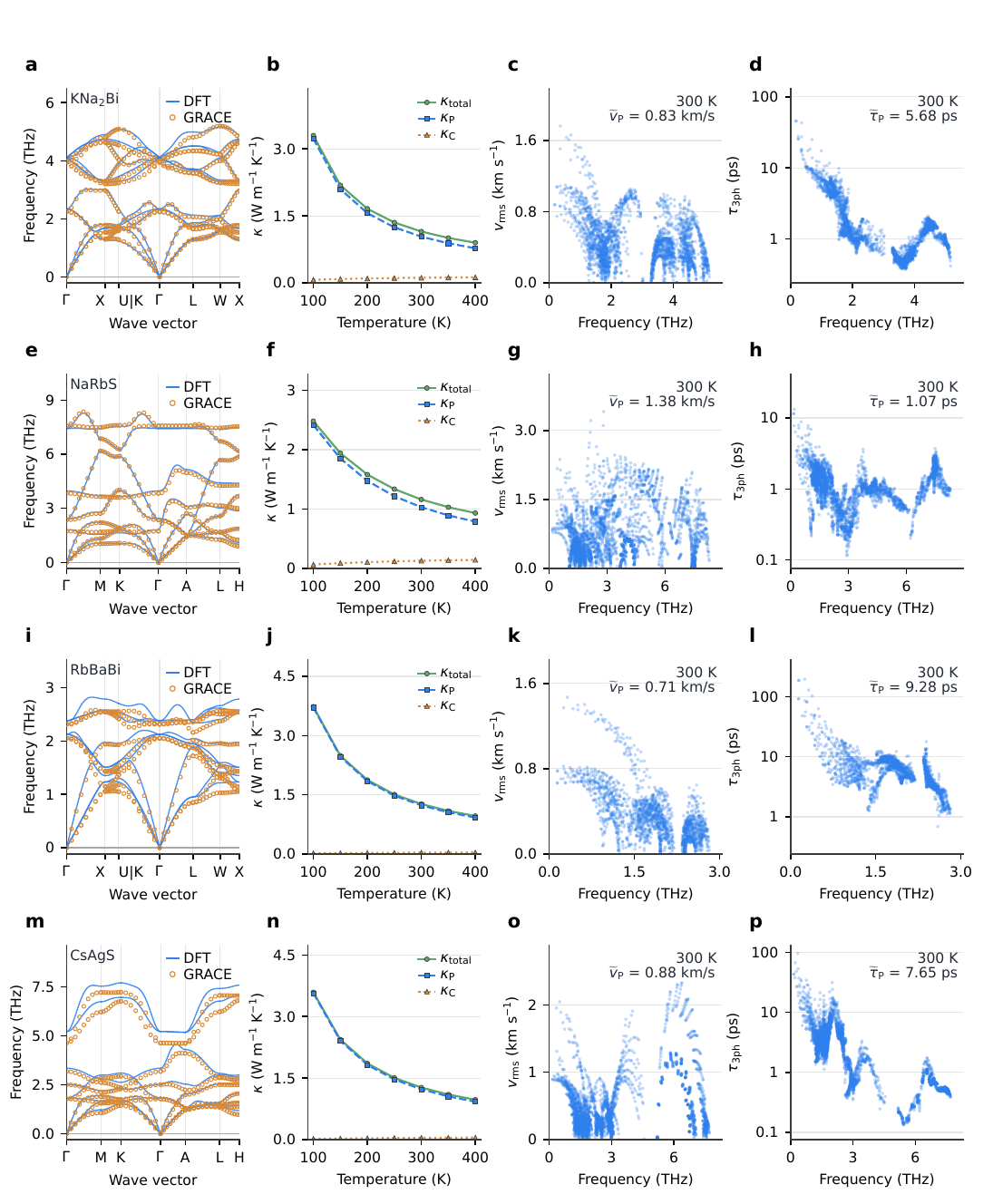}
\caption{Phonon and transport properties of $\mathrm{KNa_{2}Bi}$, $\mathrm{NaRbS}$, $\mathrm{RbBaBi}$, $\mathrm{CsAgS}$ (top to bottom). Columns show dispersions, total/population/coherence conductivity from 100--400 K, group velocity and three-phonon lifetime at 300 K. Definitions and quality markers are given above.}
\label{fig:sm-transport-13}
\end{fixedfigure}

\begin{fixedfigure}
\includegraphics[width=\textwidth,height=.88\textheight,keepaspectratio,trim=0 4pt 0 20pt,clip]{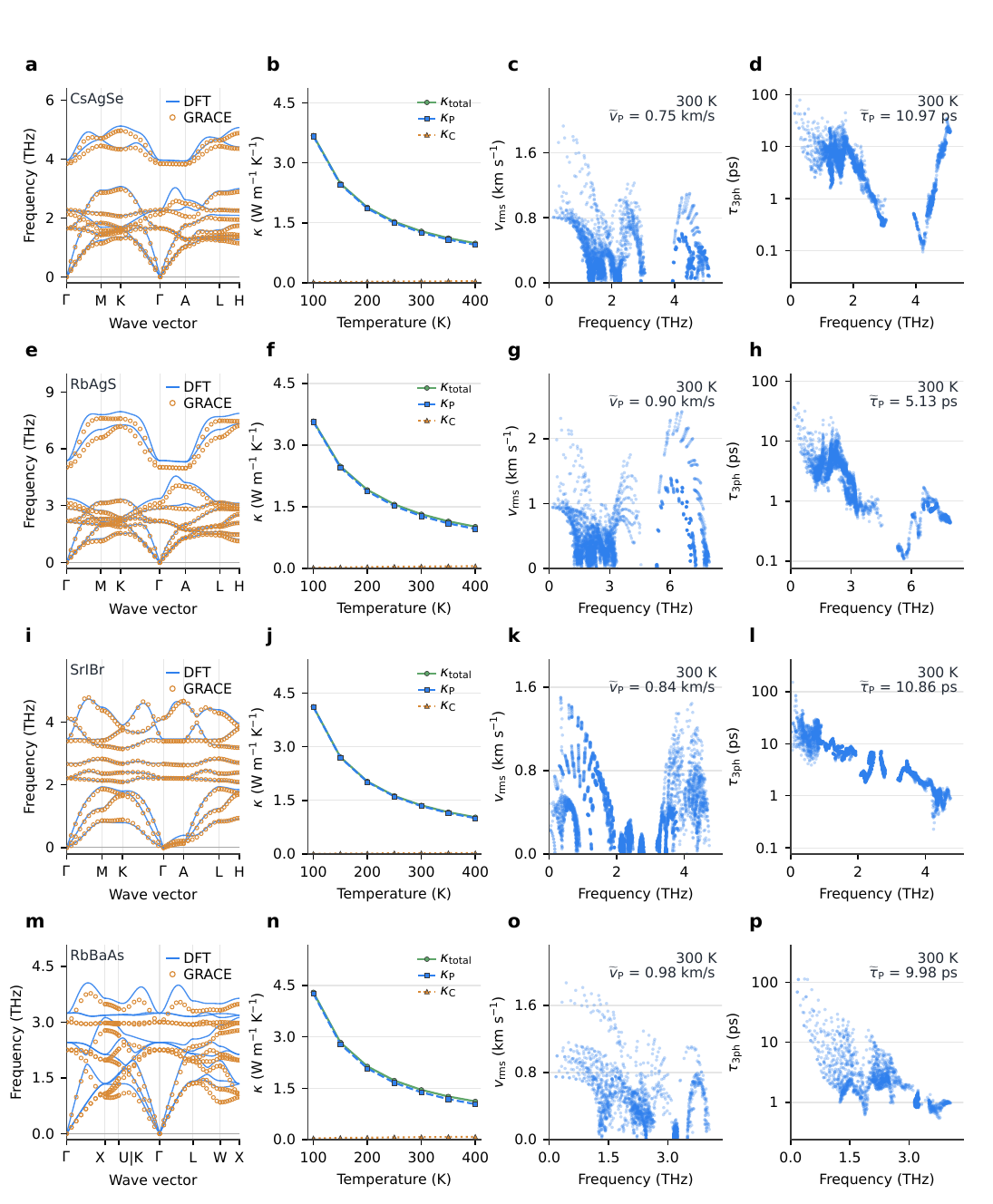}
\caption{Phonon and transport properties of $\mathrm{CsAgSe}$, $\mathrm{RbAgS}$, $\mathrm{SrIBr}$, $\mathrm{RbBaAs}$ (top to bottom). Columns show dispersions, total/population/coherence conductivity from 100--400 K, group velocity and three-phonon lifetime at 300 K. Definitions and quality markers are given above.}
\label{fig:sm-transport-14}
\end{fixedfigure}

\begin{fixedfigure}
\includegraphics[width=\textwidth,height=.88\textheight,keepaspectratio,trim=0 4pt 0 20pt,clip]{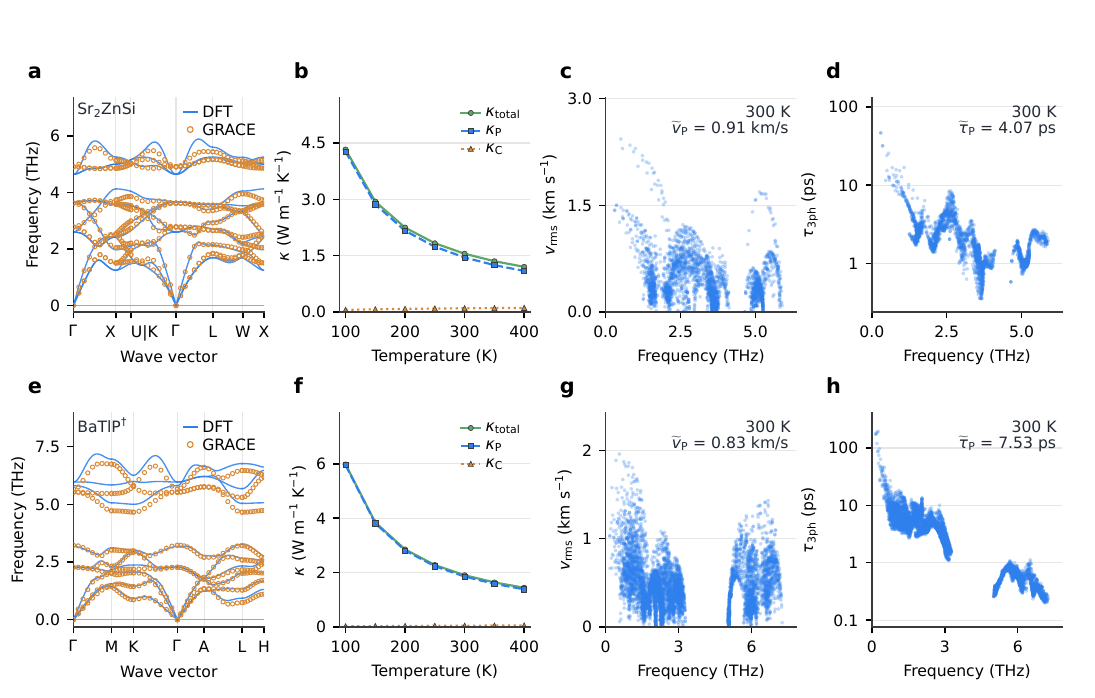}
\caption{Phonon and transport properties of $\mathrm{Sr_{2}ZnSi}$, $\mathrm{BaTlP}$ (top to bottom). Columns show dispersions, total/population/coherence conductivity from 100--400 K, group velocity and three-phonon lifetime at 300 K. Definitions and quality markers are given above.}
\label{fig:sm-transport-15}
\end{fixedfigure}

\end{document}